\documentclass[10pt]{article}
\usepackage[OE]{express}

\begin{document}

\title{Surface acoustic waves for acousto-optic modulation in buried silicon nitride waveguides.}

\author{Peter J.M. van der Slot\authormark{*}, Marco A.G. Porcel, and Klaus-J. Boller}

\address{Laser Physics and Nonlinear Optics, Mesa${+}$ Institute for Nanotechnology\\
Department for Science and Technology, University of Twente, Enschede, The Netherlands}
\email{\authormark{*}p.j.m.vanderslot@utwente.nl} 

\begin{abstract}
We theoretically investigate the use of Rayleigh surface acoustic waves (SAWs) for refractive index modulation in optical waveguides consisting of amorphous dielectrics. Considering low-loss Si$_3$N$_4$ waveguides with a standard core cross section of 4.4$\times$0.03~$\mu$m$^2$ size, buried 8-$\mu$m deep in a SiO$_2$ cladding we compare surface acoustic wave generation in various different geometries via a piezo-active, lead zirconate titanate film placed on top of the surface and driven via an interdigitized transducer (IDT). Using numerical solutions of the acoustic and optical wave equations, we determine the strain distribution of the SAW  under resonant excitation. From the overlap of the acoustic strain field with the optical mode field we calculate and maximize the attainable amplitude of index modulation in the waveguide. For the example of a near-infrared wavelength of 840~nm, a maximum shift in relative effective refractive index of 0.7x10$^{-3}$ was obtained for TE polarized light, using an IDT period of 30 - 35~$\mu$m, a film thickness of 2.5 - 3.5~$\mu$m, and an IDT voltage of 10~V. For these parameters, the resonant frequency is in the range 70 - 85~MHz. The maximum shift increases to 1.2x10$^{-3}$, with a corresponding resonant frequency of 87~MHz, when the height of the cladding above the core is reduced to 3~$\mu$m. The relative index change is about 300-times higher than in previous work based on non-resonant proximity piezo-actuation, and the modulation frequency is about 200-times higher. Exploiting the maximum relative index change of 1.2$\times$10$^{-3}$ in a low-loss balanced Mach-Zehnder modulator should allow full-contrast modulation in devices as short as 120~$\mu$m (half-wave voltage length product = 0.24~Vcm).
\end{abstract}


\bibliographystyle{osajnl}

\section{Introduction}
Integrated optical waveguides fabricated from stoichiometric silicon nitride (Si$_3$N$_4$) using low pressure chemical vapor deposition (LPCVD) offer ultra-low loss propagation and are transparent from visible to near-infrared wavelengths~\cite{bauters_ultra-low-loss_2011,worhoff_2015}. Therefore, these optical waveguides are of high interest for numerous applications. These include communications~\cite{heck_2014}, programmable quantum photonic processors~\cite{taballione_2018}, sources for nonlinear microscopy~\cite{epping_integrated_2013}, optical coherence tomography~\cite{nguyen_2012}, biosensors~\cite{zinoviev_2008}, microwave photonics~\cite{roeloffzen_2013, marpaung_integrated_2013}, ultra-narrow bandwidth hybrid lasers~\cite{oldenbeuving_2013,fan_2017} and supercontinuum generation~\cite{halir_ultrabroadband_2012,epping_-chip_2015}. Due to a high index contrast with regard to the SiO$_2$ cladding, these waveguides enable dense, complex and reconfigurable integrated photonic circuits~\cite{taballione_2018, zhuang_2014,xiong_2015,zhuang_2015}. Furthermore, on-chip adiabatic tapering allows for efficient coupling to other waveguide technologies, such as InP, silicon on oxide or fibers, in particular for light generation and detection~\cite{worhoff_2015}.

Typically, light modulation in silicon nitride waveguides relies on the thermo-optic effect and is based on a thermally induced phase shift between the two arms of a Mach-Zehnder interferometer~\cite{ovvyan_2016}. State-of-the-art thermo-optic modulators provide up to 1~kHz modulation speed, while the dissipation of heating power is often undesired, because it can be as large as 500~mW per modulator~\cite{roeloffzen_2013}. Applications that rely on a high density of modulators, \textit{e.g.}, as in reconfigurable photonic circuits~\cite{xiong_2015,taballione_2018}, would greatly benefit from modulation techniques with lower dissipation, while applications needing fast modulation of the light would greatly benefit from techniques with higher switching or modulation speeds. 

An alternative method for index modulation is using the electro-optic effect, which allows for efficient and compact modulators~\cite{tang_2005} and also allow for high modulation frequencies~\cite{tang_2005b, alexander_2018}. For example, Lithium Niobate has excellent electro-optic, nonlinear optical and piezo-electric properties \cite{tsai_guided-wave_2013} and is a common material to build discrete modulators and wavelength convertors, operating in the infrared to the visible part of the spectrum. However, to merge Lithium Niobate modulators into a Si$_3$N$_4$ photonic circuit requires heterogeneous bonding \cite{chang_2017}. To characterize electo-optic modulators, a figure-of-merit, given by the half-wave voltage length product, $V_{\pi}L$, is used. Using a Si$_3$N$_4$ strip waveguide on top of a BaTiO$_3$ thin film, Tang \textit{et al.} realized values of $V_{\pi}L = 0.25$ and 0.5~Vcm for a 5~mm long waveguide modulator and for a wavelength of 0.95~$\mu$m and 1.56~$\mu$m, respectively~\cite{tang_2005}. A 3-dB modulation bandwidth of 15~GHz was demonstrated for such a modulator using a wavelength of 1.56~$\mu$m~\cite{tang_2005b}. Alexander \textit{et al.} used a hybrid lead zirconate titanate (PZT) on Si$_3$N$_4$ waveguide to construct a C-band ring modulator with a modulation bandwidth of 33~GHz and $V_{\pi}L = 3.2$~Vcm~\cite{alexander_2018}. A drawback of these modulators based on the electro-optic effect is the increased propagation loss as the optical mode needs to overlap with the electro-optic material to obtain a strong electro-optic effect.  

A low-loss technique, which may provide both low driving power and high modulation frequency, can be based on the surface acoustic wave (SAW) induced strain-optic effect~\cite{brewster_1815,yariv_2007} where stress induced by a SAW results in a change of the effective refractive index. By using a so-called interdigital transducer (IDT), each of the individual electrodes of the transducer produce its own propagating SAW and all SAWs constructively interfere along the axis of the transducer when it is driven at the acoustic frequency of these SAWs. This enhances the response over that of a single electrode non-resonant proximity piezo-actuation of the strain~\cite{hosseini_2015,epping_2017}. The strain-optic effect has been studied in various integrated photonic systems~\cite{schriever_2012,hosseini_2015,epping_2017}. Specifically, in the waveguide platform investigated here (LPCVD Si$_3$N$_4$/SiO$_2$), and using a source for the strain that is completely outside the region where the optical mode resides, there are only two implementations investigated at two wavelengths so far, both based on the same operating principle. In~\cite{hosseini_2015}, Hosseini \textit{et al.} showed an approach with a 2~$\mu$m-thick layer of crystalline PZT deposited on top of a silicon nitride Mach-Zehnder interferometer (MZI), with the core of the waveguides positioned 8~$\mu$m below the PZT layer, for light modulation at a wavelength of 640~nm. Via an electrode placed on top of the PZT layer above one of the interferometer arms of the MZI, the stress within that arm could be locally controlled via the electrode voltage. The power consumption was reduced significantly, by six orders of magnitude, compared to thermal modulators. Also, the modulation frequency could be increased up to 600~kHz (at -3~db bandwidth). However, further increasing the modulation frequency was not possible as this would require a smaller electrode capacitance, and, consequently, a smaller electrode area will lead to a smaller induced stress and thereby a smaller induced index change. The maximum index modulation remained  rather small, at around 5 $\times$10$^{-6}$ with an optimum geometry. Consequently, the minimum $V_{\pi}L$  realized is only 12.5~Vcm. In~\cite{epping_2017}, Epping \textit{et al.} introduced a novel, low loss and ultra-low power modulator geometry, based on the same operating principle, for operation at the telecommunication C-band, requiring a half-wave driving voltage of 34~V for an optical wavelength of 1550~nm and using a modulator with a total length of 14.8~mm. The maximum modulation frequency was limited to 10~kHz.

Much higher modulation frequencies are possible with a more sophisticated electrode structure, using IDTs that resonantly excite surface acoustic waves. For instance, optical modulation at a frequency of 520~MHz was demonstrated for a compact MZI consisting of conventional ridge waveguides of GaAs with a length of the active interaction region of only 15~$\mu$m~\cite{deLima_2006}, while acousto-optic modulation of photonic resonators on thin polycrystalline aluminum nitride films have demonstrated modulation frequencies reaching well into the microwave range~\cite{tadesse_2014, tadesse_2015}.  

Here, we theoretically investigate exploiting SAW-induced effective refractive index changes for realizing high-speed, compact and low-loss modulators with Si$_3$N$_4$ waveguides. Using numerical methods we calculate the index modulation experienced by the fundamental optical mode propagating through a Si$_3$N$_4$ core buried in a SiO$_2$ cladding. The special interest in this particular geometry is that no deterioration of the ultra-low optical propagation loss is expected. The reason for that is that the cladding is taken sufficiently thick to make the optical field negligible at the location of the thin PZT film that is on top of the cladding. At the same time, the penetration depth of SAWs is large enough, on the order of the acoustic wavelength, $\lambda_{\mathrm{R}}$, in the material~\cite{rayleigh_waves_1885, lima_modulation_2005}, which allows for a good overlap of the SAW with the optical wave even for high modulation frequencies in the 100~MHz range. The SAW is considered to be launched using an IDT. Compared to the unstructured electrode arrangement used in a proximity strain-optics design~\cite{hosseini_2015,epping_2017}, we show below that the fine structuring of the IDT allows typically 200-times higher modulation frequencies while resonant excitation yields a 300-fold increase in index modulation. Another advantage of employing SAWs is that tensile strain can be applied to one interferometer arm, and, simultaneously, compressive strain can be applied to the other arm, which effectively reduces the length of the arms by a factor of two to obtain full light modulation~\cite{deLima_2006}. 

\begin{figure}[t]
	\centering
	\includegraphics[width=0.85\linewidth]{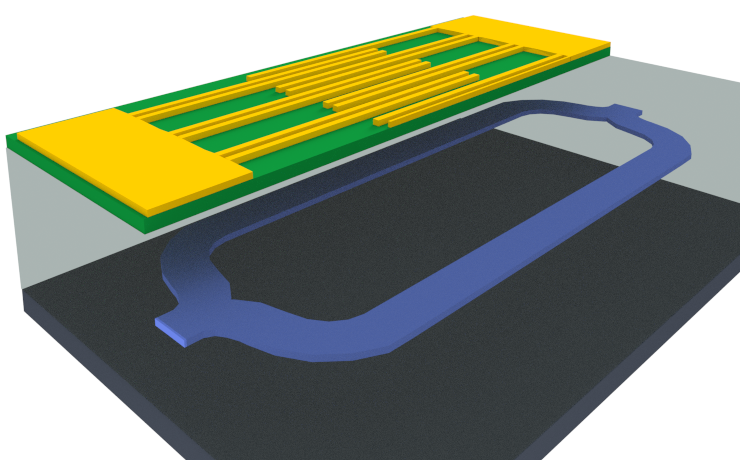}
	\caption{\label{fig:shg_mzi}
		Schematic view of a waveguide (in dark blue) based Mach-Zehnder interferometer. Half of the electrode structure  (in yellow) and piezo layer (in green) is not shown to enable a full view on the interferometer (dark blue waveguides). The interferometer waveguides are buried in SiO$_{2}$ (light gray) deposited on a silicon substrate (dark Grey).}
\end{figure}

In the following we consider acousto-optic modulation using a MZI, in a setting where the acoustic wave propagates perpendicular to the optical waveguide axis of the two arms of the MZI as shown in Fig.~\ref{fig:shg_mzi}. We briefly discuss the relation between strain and the refractive index and the surface acoustic wave of interest.  We then present the geometry studied and how the simulations are performed. We investigate how the induced strain and, consequently, the effective refractive index, depends on the thickness of the PZT layer and the period of the IDT used to generated the SAW. Finally, we optimize the location of the core and we use the maximum change in effective refractive index to determine the required length of the MZI to obtain full modulation of the optical wave. 
The optimized maximum relative change of the effective waveguide index in our arrangement was found to be 0.12 \%, at a frequency of 87~MHz when using a voltage amplitude of 10~V. Such a change in effective refractive index yields full modulation with a relatively short arm length of 120~$\mu$m ($V_{\pi}L=0.24$~Vcm). This is about 100-times shorter than with the proximity piezo method described above.

\section{Acousto-optic refractive index modulation}
The response of a material to an applied electric field, an acoustic wave, or a combination of the two, strongly depends on the type of material. Here, we are interested on one hand in generating a strong acoustic wave using the piezoelectric effect~\cite{kay_1955, sundar_1992}. On the other hand, we intend to use strain in amorphous waveguide core ($\text{Si}_3\text{N}_4$) and cladding (SiO$_2$) materials (which do not possess a piezoelectric effect) to cause strain-induced changes in refractive index~\cite{kay_1955, sundar_1992}. The strain can be compressive or tensile, leading to an increase or decrease in local refractive index, respectively. In a microscopic picture, strain changes both the number of microscopic dipoles per unit volume and the microscopic potential. The volume change modifies the induced dipole driven by the applied optical field and thus changes the optical susceptibility tensor, $\boldsymbol{\chi}$, of the material. In the most general case of an anisotropic material, the relation between the change in the inverse  of the dielectric tensor and the applied strain is given by~\cite{yariv_2007,boyd_nonlinear_2013},

\begin{equation}\label{eq:tensorrelation}
\left[ \Delta \left( \mathbf{\varepsilon}^{-1} \right) \right]_{ij} = (\mathbf{pS})_{ij} = \sum_{kl} \mathbf{p}_{ijkl} \mathbf{S}_{kl} , 
\end{equation}

where $\boldsymbol{\varepsilon}=\varepsilon_{0}(\mathbf{1}+\boldsymbol{\chi})$ is the dielectric tensor, $\varepsilon_{0}$ is the vacuum permittivity,  $\mathbf{S}$ is the strain tensor and $\mathbf{p}$ is the dimensionless strain-optic tensor. The indices, $i, j, k$ and $l$ designate the three Cartesian coordinates.  For isotropic materials, such as the amorphous materials investigated here, and assuming small changes in the inverse dielectric tensor, Eq. (\ref{eq:tensorrelation}) can be simplified. In this case, the related change in refractive index is given by~\cite{borrelli_1968,korpel_1996}
\begin{equation}\label{eq.delta_nx}
\Delta n_{x} = -\frac{1}{2} n^{3}_{0} ( p_{11} S_{x} + p_{12} S_{y} ).
\end{equation}
Here, $\Delta n_{x}$ is the change in refractive index for light polarized linearly along the $x$ direction, $n_{0}$ is the refractive index of the material in absence of any strain, $S_{i}$ is the strain applied in the $i$-direction ($i=x,y$) and the contracted indices notation is used~\cite{yariv_2007}. The change in refractive index for the other polarization direction is obtained via exchanging the strain-tensor components,
\begin{equation}\label{eq.delta_ny}
\Delta n_{y} = -\frac{1}{2} n^{3}_{0} ( p_{12} S_{x} + p_{11} S_{y} ).
\end{equation}

In this work, strain in the region of the optical mode is induced by an acoustic wave. As the optical mode is confined to an area just a few micrometer below the surface of the cladding, for obtaining a strong interaction, the surface acoustic wave (SAW) is the most appropriate acoustic wave to consider for a strong interaction. The reason is that a SAW travels along the surface of an elastic material and most of its energy and strain are confined within a small region, with a thickness of the order of the acoustic wavelength, below the surface and thus can provide good overlap between the strain induced by the SAW and the optical mode. It should be noted that in general various types of acoustic waves can be excited \cite{rebiere_1992,arnau_2008}. However, by taking the right crystal orientation for the piezoelectric material and an appropriate design for the IDT transducer, the different types of acoustic waves each have their own distinct frequency and include Rayleigh SAWs. Therefore, by finding and applying the right excitation frequency we can selectively excite the Rayleigh SAW wave, which is characterized by a correlated transverse and longitudinal motion at the surface. This results in  volume elements traversing an elliptical path when the wave passes~\cite{telford_1990}. The motivation to investigate Rayleigh waves is that the considered SAW (SAW refers to a Rayleigh surface acoustic wave in the remainder of this work) has low dispersion, as long as the elastic modulus near the surface does not change~\cite{telford_1990}, making them suitable for the modulation of broadband signals~\cite{hashimoto_2000}. 

If the SAW is launched perpendicular to the two arms of a MZI as depicted in Figs.~\ref{fig:shg_mzi} and \ref{fig:geometry}, and the two arms of length $L$ are separated by half an acoustic wavelength, one arm experiences compressive strain and the other tensile strain. The phase shift of light leaving either interferometer arm with respect to the light entering the interferometer is given by
\begin{equation} \label{eq.delta.phi}
\Delta \varphi = 2\pi \Delta n \frac{L}{\lambda},
\end{equation}
where $\Delta n$ is the change in effective refractive index, $n_{\mathrm{eff}}$, of the optical mode, which is opposite in sign for the two arms, $n_{\mathrm{eff}}=\frac{c\beta}{\omega}$, $c$ is the speed of light in vacuum, $\beta$ is the propagation constant for the fundamental mode, $\omega$ is the light frequency and $\lambda$ is the vacuum wavelength. When the light is combined at the output of the MZI, the total phase difference is 
\begin{equation} \label{eq.delta.phitot}
\Delta \varphi_{t} = 4\pi |\Delta n| \frac{L}{\lambda}.
\end{equation}
A total phase difference of $\Delta \varphi_{t}=\pi$ is required for full light modulation. Therefore, when the length $L$ is equal to 
\begin{equation} \label{eq.length}
L = \frac{\lambda}{4|\Delta n|},
\end{equation}
light is fully modulated with a modulation frequency equal to the frequency of the SAW. The latter equation shows that a weak index modulation (small $\Delta n$) would require long arm lengths or interaction lengths, which is undesired for compact, integrated waveguide circuits with a high density of components.

For providing full modulation also with short interaction lengths, a SAW is ideally created in a material having a large piezoelectric coefficient. For our calculations we consider lead zirconate titanate (PZT), because PZT is known as a high performance piezoelectric material  and commonly used in actuators and sensors~\cite{saito_2004}. Another advantage is that thin PZT layers have already been successfully deposited on silicon nitride waveguides~\cite{hosseini_2015,epping_2017,alexander_2018}. Both a Ti/Pt bilayer ~\cite{hosseini_2015,epping_2017}, and a lanthanide- based layer~\cite{alexander_2018} have been used to epitaxially grow PZT on top of the amorphous cladding. 
The PZT layer itself may be grown using pulsed laser deposition~\cite{hosseini_2015} or alternative techniques like liquid-phase growth~\cite{fukushima_1984}.  The different seeding techniques allow configurations with and without a conducting layer between the amorphous cladding and crystalline PZT. The generation and comparison of SAWs produced by various IDT configurations is presented in the next sections.

\section{Geometry and simulation domain}
\begin{figure}[t]
	\centering
	\includegraphics[width=0.75\linewidth]{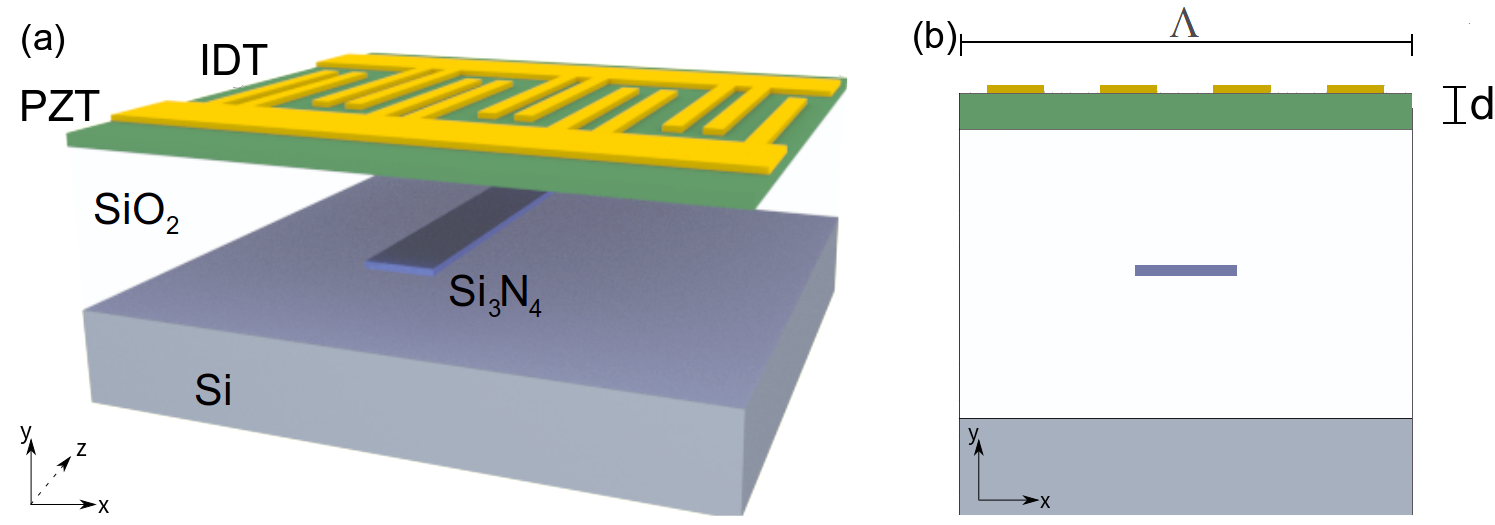}
	\includegraphics[width=\linewidth]{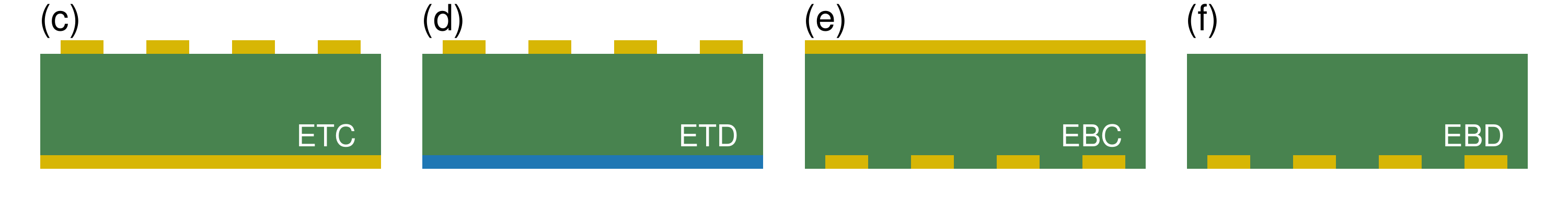}
	\caption{(a) Schematic view of the geometry comprising a Si$_3$N$_4$ core (4.4x0.03~$\mu$m$^{2}$, dark blue) centered in a 16-$\mu$m thick SiO$_2$ cladding (light gray) on a Si substrate (dark gray). A crystalline PZT layer on top of the cladding (green) in combination with an IDT (yellow) is used for exciting surface acoustic waves (SAWs). (b) The cross-section shows the corresponding two-dimensional unit cell across a single period $\Lambda$ of the IDT. The thickness of the Si layer included in the unit cell is proportional to $\Lambda$ to ensure a negligible SAW amplitude at the lower boundary of the unit cell. The layer between the PZT and the SiO$_{2}$ layers is either conductive (yellow) or dielectric (light blue) and functions as seed layer for the crystalline growth of the PZT layer on top of the amorphous SiO$_{2}$ or IDT electrode. (c)-(f) Illustrate the various combinations of IDT locations with and without opposite conductive layer, following the same color coding. (c) \texttt{\footnotesize ETC}, (d) \texttt{\footnotesize ETD},(e) \texttt{\footnotesize EBC},(f) \texttt{\footnotesize EBD}. Features are not to scale.}
	\label{fig:geometry}
\end{figure}

The geometries considered here are shown schematically in Fig.~\ref{fig:geometry}(a) and as cross-sections in Figs.~\ref{fig:geometry}(b)-\ref{fig:geometry}(f). The optical design comprises a typical low-loss optical waveguide having a $\text{Si}_3\text{N}_4$ core of height 30~nm and width 4.4~$\mu$m embedded symmetrically in a 16~$\mu$m thick SiO$_2$ cladding on top of a Si wafer substrate, \textit{i.e.}, at 8~$\mu$m distance above the Si-SiO$_2$ interface. The strain-optic tensor is not known for Si$_3$N$_4$. However, due to the small core thickness the optical mode is mostly outside the core and a modulation of the effective refractive index is dominantly caused by the strain-optic effect in the SiO$_2$ cladding. The thickness of the cladding, 16~$\mu$m, is large enough to ensure that the optical mode is completely confined to the core and cladding, in order to render optical losses due to surface layers or the substrate negligible. The top of the SiO$_2$ cladding contains a thin conductive or dielectric seeding layer that allows the growth of c-oriented PZT. The thickness, $d$, of the PZT layer was varied to determine the optimum thickness for excitation of the SAW. A split-finger IDT configuration is used to excite the SAW without first order Bragg reflections~\cite{hashimoto_2000}. In order to maximize the optical modulation amplitude, we investigate the effect of the location of the conductive (\textit{i.e.} gold) electrodes of the IDT and a seeding layer in four different configurations. For the convenience of the reader we identify each configuration with a three-letter abbreviation, the first two letters are \texttt{\footnotesize ET} or \texttt{\footnotesize EB} and indicate if the location of the IDT electrode is at the top or bottom of the PZT film, respectively. The last letter is either \texttt{\footnotesize C} or \texttt{\footnotesize D} and indicates that the terminating layer on the opposite side of the PZT film is conductive or dielectric, respectively. The four configurations \texttt{\footnotesize ETC}, \texttt{\footnotesize ETD}, \texttt{\footnotesize EBC} and \texttt{\footnotesize EBD} are schematically shown in Figs.~\ref{fig:geometry}(c)-\ref{fig:geometry}(f), respectively. Configurations \texttt{\footnotesize EBC} and \texttt{\footnotesize EBD} have both a thin seed dielectric nanosheet deposited on top of the IDT electrode and remainder of the cladding to allow crystalline growth of the PZT layer. Because the dielectric seed layer is only a few nm thick and is considered to have perfect adhesion to both materials, it would not notably affect the acoustic wave and is not included in the model.

The IDT generates a SAW and the associated strain induces a change in refractive index in both the cladding and the core. However, the optical modulation amplitude is expected to depend on the strain distribution of the excited SAW and its overlap with the optical mode. In order to determine the degree of overlap, the strain distribution is calculated by finding the fundamental SAW eigenmode for the four configurations using a finite-element eigenmode solver~\cite{comsol_2014}. For these calculations, we use the two-dimensional unit cell shown in Fig. \ref{fig:geometry}(b), which consists of one period of the split finger IDT electrode and the layers below (and layers above in case of configurations \texttt{\footnotesize EBC} and \texttt{\footnotesize EBD}). To maximize the strain at the location of the optical mode, the waveguide core is positioned in the horizontal direction, $x$ in Fig.~\ref{fig:geometry}, to lie symmetrically underneath the gap between the two IDT electrodes. Platinum is selected for the conductive seed layer material (gray) and gold (yellow) for the IDT electrodes and the conductive layer on top of the PZT. Both layers as well as the gold electrodes of the IDT are taken as 100~nm thick. 

The acoustic boundary conditions applied to the unit cell are a free displacement condition at the PZT-air interface, a zero displacement condition at the bottom of the Si substrate and a periodic boundary condition at the two remaining boundaries. In order to ensure that the SAW has negligible amplitude near the bottom of the substrate and the zero displacement boundary condition does not affect the solution, the height of the Si substrate was found not to affect the solutions for heights larger than 5$\Lambda$, where $\Lambda$ is the period of the IDT electrode. 

For the calculation of the optical field distribution, we use an optical eigenmode solver \cite{comsol_2014} on the domain consisting of the core and cladding with  zero-field as optical boundary condition at all outer boundaries of this domain. This is well-justified because with the given index and size parameters the optical field is confined closely around the core as compared to the thickness of the cladding. The refractive indices for the Si$_3$N$_4$ core and SiO$_2$ cladding materials are taken from Luke \textit{et al.}~\cite{luke_broadband_2015}. 

To calculate the resonant acoustic frequencies for each of the configurations as a function of both the thickness, $d$, of the PZT layer and period, $\Lambda$, of the IDT electrode, we use an eigenmode solver. Note that the resonant frequency corresponds to the modulation frequency of the light in a properly configured Mach-Zehnder interferometer. To determine the effective refractive index at the location of the optical mode, it is required that the applied voltage is chosen to oscillate at the resonant frequency. The electro-mechanical coupling coefficient was determined by performing a frequency-domain simulation using the same unit cell as shown in Fig.~\ref{fig:geometry}(b) and calculating the strain distribution when a sinusoidal voltage with a given amplitude (we chose 10~V) and a frequency equal to the resonant acoustic frequency is applied to the IDT electrode. As we include acoustic damping, the calculated SAW response corresponds to a transducer with a length larger than the propagation loss length of the SAW. Also for this study, we investigated the dependence of the induced strain on $d$ and $\Lambda$. The various isotropic material properties used in the simulation are listed in Table~\ref{tab:materials}~\cite{lazan_damping_1968, zhang_2014, comsol_2014, morgan}. The elasticity matrix $C$ used for PZT and silicon in the simulation is given by~\cite{zhang_2014}
\begin{equation}
C= \begin{pmatrix}
c_{11} & c_{12} & c_{21} & 0 & 0 & 0\\
c_{12} & c_{11} & c_{21} & 0 & 0 & 0\\
c_{21} & c_{21} & c_{33} & 0 & 0 & 0\\
0 & 0 & 0 & c_{44} & 0 & 0 \\
0 & 0 & 0 & 0 & c_{44} & 0 \\
0 & 0 & 0 & 0 & 0 & c_{66}
\end{pmatrix},
\end{equation}
where the coefficients for PZT are $c_{11}=134.87$~GPa, $c_{12}=67.89$~GPa, $c_{21}=68.08$~GPa, $c_{33}=113.30$~GPa, $c_{44}=22.22$~GPa and $c_{66}=33.44$~GPa. For silicon there are only three independent coefficients $c_{11}=c_{33}=166$~GPa, $c_{12}=c_{21}= 64$~GPa, and $c_{44}=c_{66}=80$~GPa.
The piezoelectric coupling tensor, $\boldsymbol{d}$, is given by~\cite{morgan}
\begin{equation}
\mathbf{d}= \begin{pmatrix}
0 & 0 & 0 & 0 & 440 & 0 \\
0 & 0 & 0 & 440 & 0 & 0 \\
-60 & -60 & 152 & 0 & 0 & 0
\end{pmatrix} \times 10^{-12} \frac{\mathrm{C}}{\mathrm{N}},
\end{equation}
and the relative permittivity tensor, $\boldsymbol{\varepsilon}^{T}$, is given by~\cite{morgan}
\begin{equation}
\mathbf{\varepsilon}^{T}= \begin{pmatrix}
990 & 0 & 0 \\
0 & 990 & 0 \\
0 & 0 & 450
\end{pmatrix}, 
\end{equation}
where the superscript $T$ indicates that the relative permittivity tensor is measured under constant stress.

\begin{table}[tb]
	\centering
	\caption{Material constants used in calculating the SAW properties. Parameters not included in the model are denoted as (\texttt{\footnotesize NI}). Parameters included in tensor form are shown as (-). }
	\label{tab:materials}
	\begin{tabular}{|| c c c c c ||}
		\hline
		\multicolumn{5}{|c|}{Material constants}\\
		\hline
		& Young's Modulus	& Poisson's Ratio 	& Density  & Damping\\ 
		& GPa             	&                             &kg/m$^3$& 	factor, $\eta$   \\
		\hline
		Pt			&	168	&	0.38	&	21450 & \texttt{\footnotesize NI}	 \\
		Au			&	70	&	0.44	&	19300 &  0.0003	\\
		PZT   		& 	-	&	-	&	7600   & 0.03\\
		SiO$_2$	& 	70	&	0.17	&	2650   & 0.0004	\\
		Si$_3$N$_4$& 	250	&	0.23	&	3100   & \texttt{\footnotesize NI}	\\
		Si  			&	-	&	-	&	2330   & \texttt{\footnotesize NI}	\\ 
		\hline
	\end{tabular}
	
\end{table}

\section{Results}
\subsection{Acoustic wave generation}
We are interested in MZI-based modulation with a maximum optical phase change in the interferometer arms. Therefore we consider a geometry where the SAW propagation direction is perpendicular to the optical axis of the waveguide, as in Figs.~\ref{fig:shg_mzi} and \ref{fig:geometry}. The modulation frequency is taken as equal to the resonant frequency of the fundamental SAW.  In this case, the acoustic wavelength, $\lambda_{\mathrm{R}}$, and frequency, $f_{\mathrm{R}}$, are equal to the period, $\Lambda$, and driving frequency of the IDT, respectively. The relation between $f_{\mathrm{R}}$ and $\lambda_{\mathrm{R}}=\Lambda$ is given by
\begin{equation}
f_{\mathrm{R}} = v_{\mathrm{R}}/\Lambda,
\label{eq:f_r}
\end{equation} 
where $v_{\mathrm{R}}$ is the phase velocity of the SAW. In general, $v_{\mathrm{R}}$ varies with $\Lambda$, $d$, the choice of materials and the different IDT configurations such that the modulation frequency of the optical wave will also vary with these parameters.

\begin{figure}[b]
	\centering
	\includegraphics[width=0.45\linewidth]{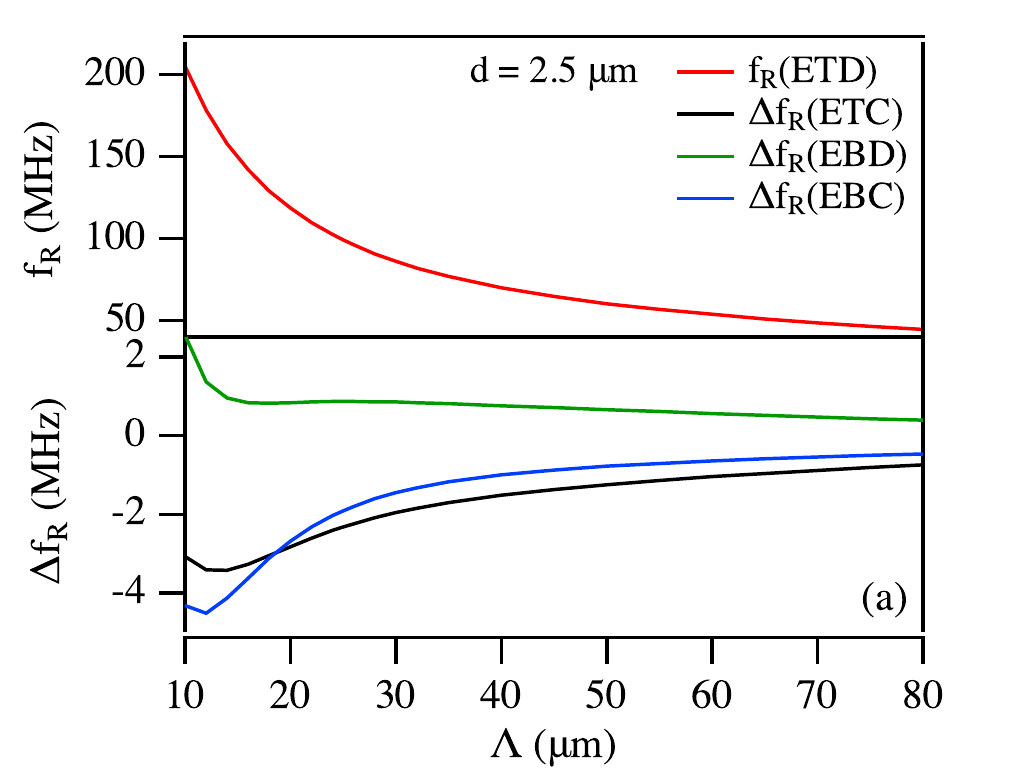}
    \includegraphics[width=0.45\linewidth]{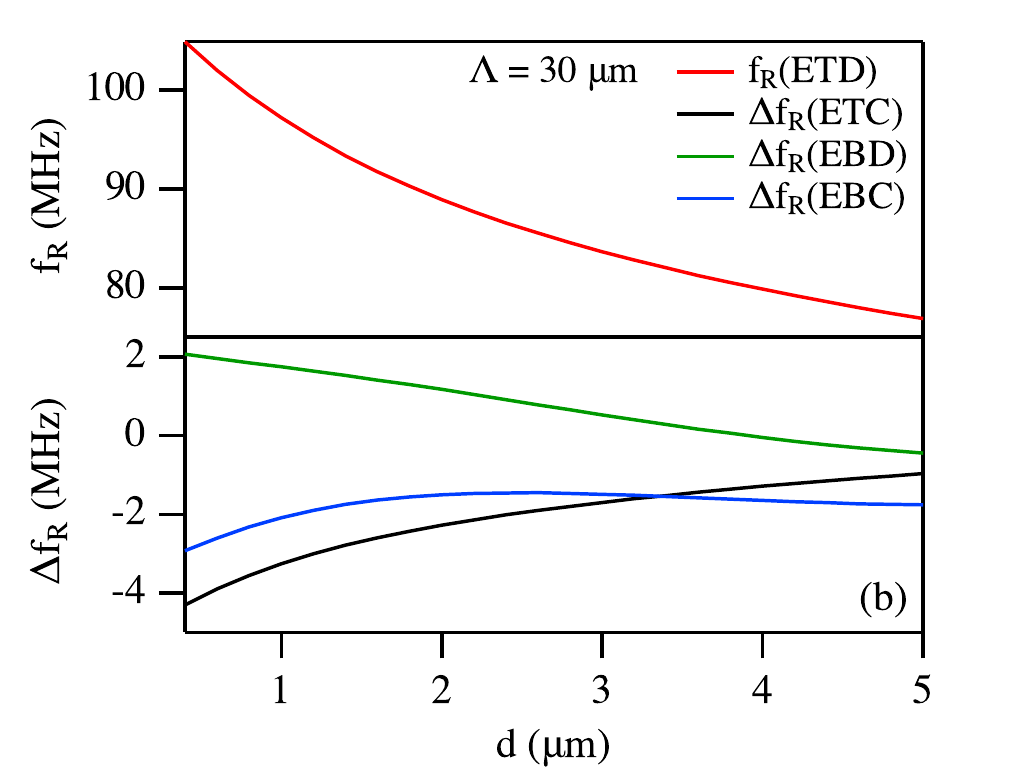}
	\caption{\label{fig:f_r}
		Resonant acoustic frequency, $f_{\mathrm{R}}$, of the fundamental Rayleigh wave for the IDT configuration \texttt{\footnotesize ETD}, and, for clarity, the frequency difference $\Delta f_{\mathrm{R}}$ with regard to the reference IDT configuration \texttt{\footnotesize ETD}, for the remaining IDT configurations \texttt{\footnotesize EBD}, \texttt{\footnotesize ETC} and \texttt{\footnotesize EBC}.The frequency is displayed as function of the period of the acoustic wave, which is equal to the period $\Lambda$ of the IDT for a fixed PZT thickness $d=2$~$\mu$m \textit(a), and as a function of the thickness $d$ of the PZT layer for an IDT period of $\Lambda=30$~$\mu$m \textit(b).} 
\end{figure}

Figure \ref{fig:f_r}(a) shows the acoustic frequency, $f_{\mathrm{R}}$, of the fundamental SAW as a function of the IDT period, $\Lambda$, for a fixed PZT layer thickness of $d=2.5$~$\mu$m for the reference geometry \texttt{\footnotesize ETD} and, for clarity, the frequency difference $\Delta f_{\mathrm{R}}(\mathrm{geom}) = f_{\mathrm{R}}(\mathrm{geom}) - f_{\mathrm{R}}(\texttt{\footnotesize ETD})$ for the three remaining geometries, where geom is equal to \texttt{\footnotesize ETC}, \texttt{\footnotesize EBD} or \texttt{\footnotesize EBC}. Similarly, Fig.~\ref{fig:f_r}(b) shows $f_{\mathrm{R}}$ for \texttt{\footnotesize ETD} and $\Delta f_{\mathrm{R}}$ for the configurations \texttt{\footnotesize ETC}, \texttt{\footnotesize EBD} and \texttt{\footnotesize EBC} as a function of the PZT thickness, $d$, for a fixed IDT period of $\Lambda=30$~$\mu$m.

Figure \ref{fig:f_r}(a) shows that the acoustic frequency monotonically decreases, approximately inversely with $\Lambda$, for all configurations and that an IDT period of  $\Lambda \sim 25$~$\mu$m is needed for a resonant acoustic frequency of around 100~MHz. Furthermore, this figure  shows that the four configurations have almost the same acoustic frequency. The largest difference between the acoustic frequencies is found for the shorter IDT periods. For $\Lambda \gtrsim 20$~$\mu$m, having the IDT at the PZT-SiO$_{2}$ interface increases the acoustic frequency somewhat, while the presence of a conductive layer reduces the acoustic frequency. Figure \ref{fig:f_r}(a) shows that for $\Lambda \gtrsim 20$~$\mu$m the reduction in resonant frequency due to conductive layer is larger than the increase caused by having the IDT at the PZT-SiO$_{2}$ interface. 

Figure \ref{fig:f_r}(b) shows that the frequency also monotonically decreases for all configurations as the layer thickness increases and that $|\Delta f_{\mathrm{R}}|$ gets smaller when $d$ increase, at least for $d \lesssim 4$~$\mu$m. Moving the IDT to the Si-SIO$_{2}$ interface (configuration \texttt{\footnotesize EBD}) results in a  somewhat larger acoustic frequency , however for $d \gtrsim 4$~$\mu$m its acoustics frequency drops below that of the reference configuration \texttt{\footnotesize ETD}. Adding a conductive layer to the configuration lowers the acoustic frequency, albeit that this effect reduces when the layer thickness increase (configurations \texttt{\footnotesize ETC}  and \texttt{\footnotesize EBC}). For small layer thickness, the resonant acoustic frequency for configuration \texttt{\footnotesize ETC} is lower than that produced by configuration \texttt{\footnotesize EBC}, while for larger layer thickness, the situation is reversed. We observe that for $d \gtrsim 4$~$\mu$m the reference configuration \texttt{\footnotesize ETD} has the highest resonant acoustic frequency.

The different resonant acoustic frequencies, found when the thickness of the layer is varied at constant IDT period, indicate that the sound velocity of the acoustic wave is affected by the amount of PZT material present. On the other hand, for a fixed geometry and varying only the IDT period, \textit{i.e}., the period of the acoustic wave, we observe a strong increase in the resonant frequency when the period decreases (Fig.~\ref{fig:f_r}(a)), as expected from Eq.~(\ref{eq:f_r}). 
In summary, for the parameters investigated and using the configuration \texttt{\footnotesize ETD} as a reference, terminating the PZT layer with a conductive layer opposite to the IDT electrode reduces the resonant acoustic frequency somewhat  and placing the IDT electrode at the PZT-SiO$_{2}$ interface increases the resonant acoustic frequency a little.

In order to determine the change in effective refractive index of the fundamental optical mode that is induced by the SAW, the strain distribution generated by the SAW within the volume of the optical mode has to be calculated. A frequency domain analysis is performed to calculate the induced strain when a sinusoidal voltage oscillating at the resonant acoustic frequency is applied to the IDT electrode. Adding acoustic damping to the cladding and piezo regions (see parameters in Table~\ref{tab:materials}), provides a physically realistic and numerically stable response. We observe that the strain found by frequency domain solver \cite{comsol_2014} is 90 degrees out of phase with the applied voltage, \textit{i.e.}, there is a time delay between strain distribution and applied voltage. For a given amplitude of the oscillating IDT voltage, we determine the strain distribution with maximum amplitude, while preserving the relative sign of the $S_{x}$ and $S_{y}$ components, and use this distribution in Eqs. (\ref{eq.delta_nx}) and (\ref{eq.delta_ny}).  

\begin{figure}[bt]
	\centering
	\includegraphics[width=0.45\linewidth]{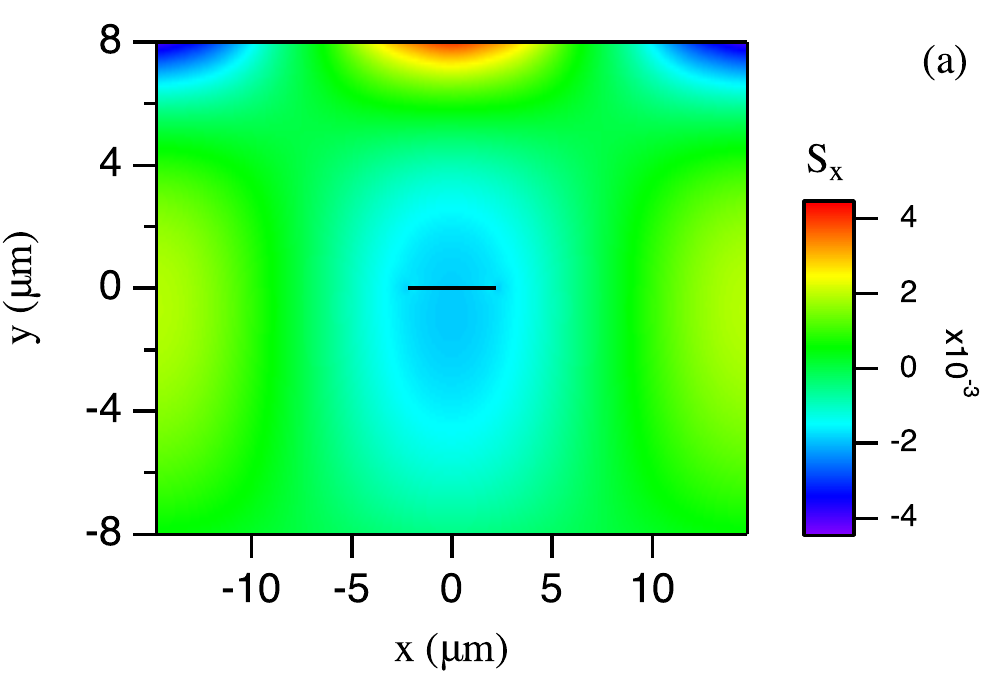}
    \includegraphics[width=0.45\linewidth]{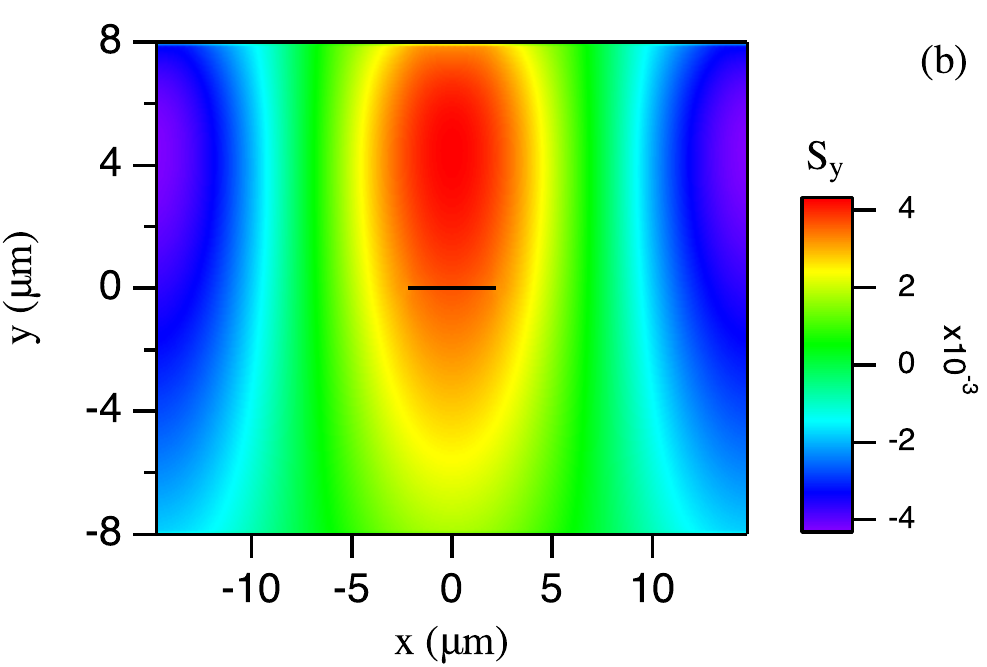}
	\caption{Strain distribution in the horizontal, $S_x$, (a) and vertical, $S_y$, (b) directions as generated by the fundamental Rayleigh wave when $\Lambda=30$~$\mu$m and $d=2$~$\mu$m. The Si$_3$N$_4$ waveguide core is shown in black at scale.} 
	\label{fig:strain}
\end{figure}

A typical example of the strain distribution generated by the fundamental SAW when a voltage signal with an amplitude of 10~V is applied to the IDT electrode is shown in Fig.~\ref{fig:strain} for the configuration \texttt{\footnotesize ETD} with $\Lambda=30$~$\mu$m and $d=2$~$\mu$m, which corresponds to modulation with $f_{\mathrm{R}}=89$~MHz. In this figure, only the region of interest is shown, \textit{i.e.}, the region containing the waveguide's core and cladding. The origin of the coordinate system coincides with the center of the optical waveguide, and $y=\pm8$~$\mu$m coincides with the PZT-SiO$_2$ and SiO$_{2}$-Si interface, respectively. The $z$-axis (along which the optical mode propagates) points along the axis of the waveguide, normal to $x$ and $y$. Figure~\ref{fig:strain}(a) shows the induced strain in the $x$-direction, $S_{x}$, and Fig.~\ref{fig:strain}(b) shows the strain in the $y$-direction, $S_{y}$. The strain distributions for the other three IDT configurations are very similar to that shown in Fig.~\ref{fig:strain}. This typical example shows that the SAW-induced strain easily extends to the core of the optical waveguide (shown as black  line in Figs.~\ref{fig:strain}(a) and \ref{fig:strain}(b)) and, therefore, a good overlap between the induced strain and optical mode is expected. However, comparing Fig.~\ref{fig:strain}(a) with Fig.~\ref{fig:strain}(b) shows that the two strain components have opposite sign at the location of the core for $\Lambda = 30$~$\mu$m and $d=2.0$~$\mu$m. Equations (\ref{eq.delta_nx}) and (\ref{eq.delta_ny}) show that, for this case, the contribution of the two strain components to the refractive index partly neutralize each other. Due to different weighting by the strain-optic coefficients, we expect different performance for TE and TM polarized optical modes, \textit{i.e.}, for optical modes with linear polarization along the $x$ and $y$ direction, respectively. In the next section we will discuss the strain produced by the SAW in more detail.  

\subsubsection*{Maximizing the strain}
\begin{figure}[hbt]
	\begin{tabular} {c c}
	\includegraphics[width=0.45\linewidth]{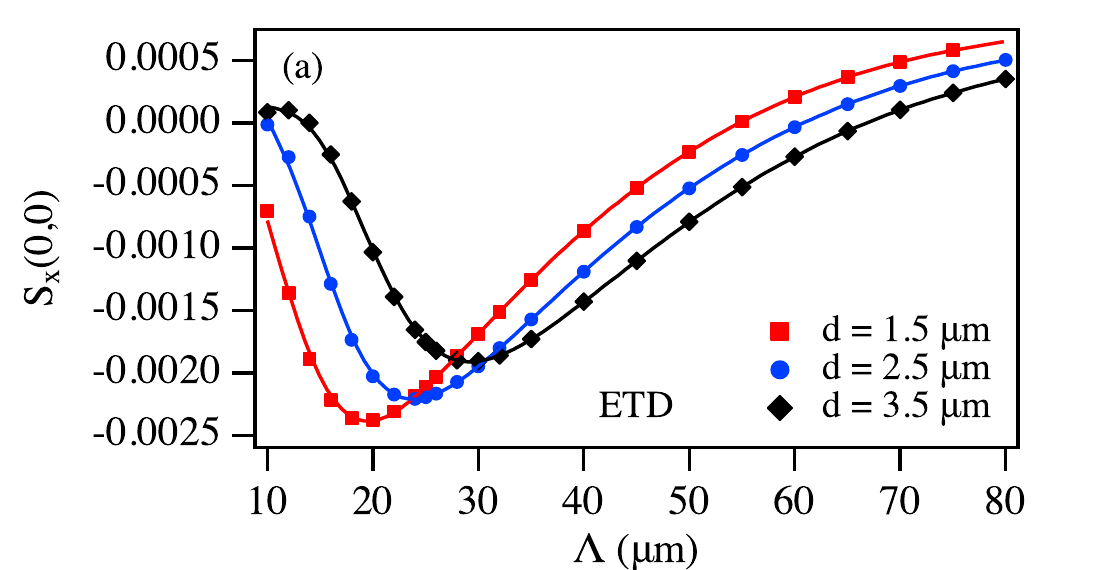} &
    \includegraphics[width=0.45\linewidth]{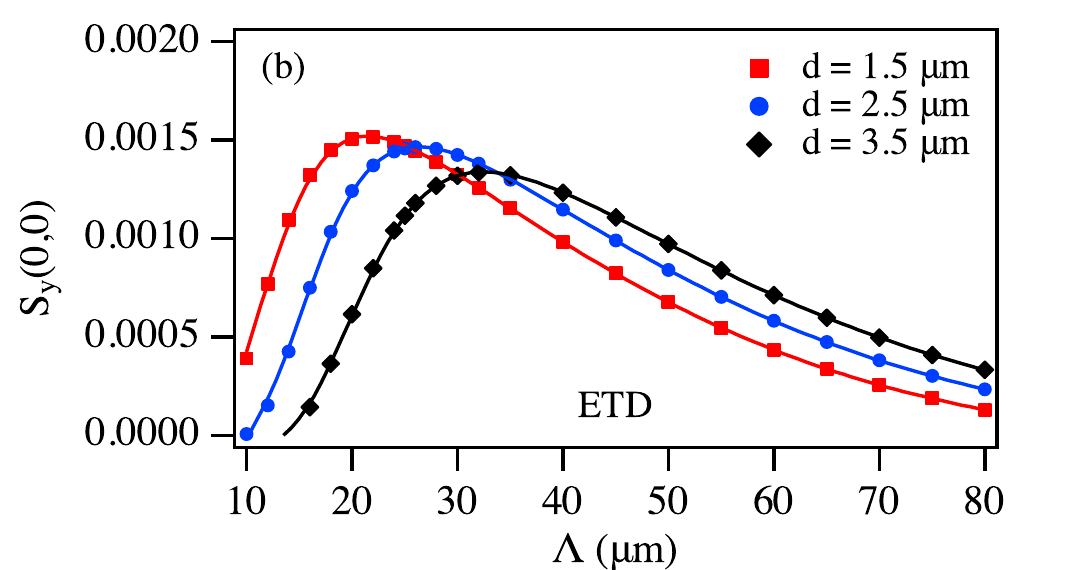}\\
    \includegraphics[width=0.45\linewidth]{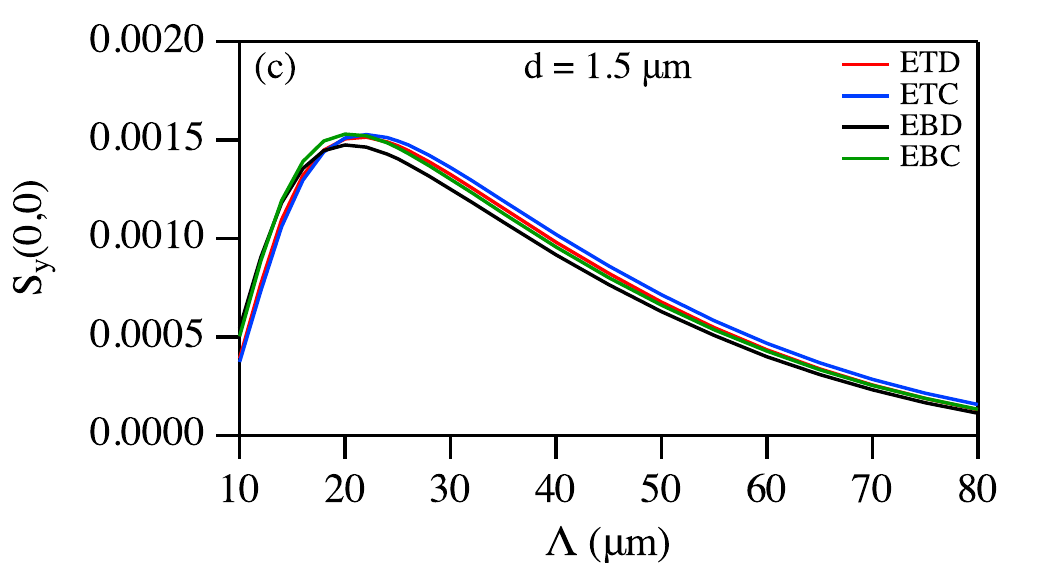} &
    \includegraphics[width=0.45\linewidth]{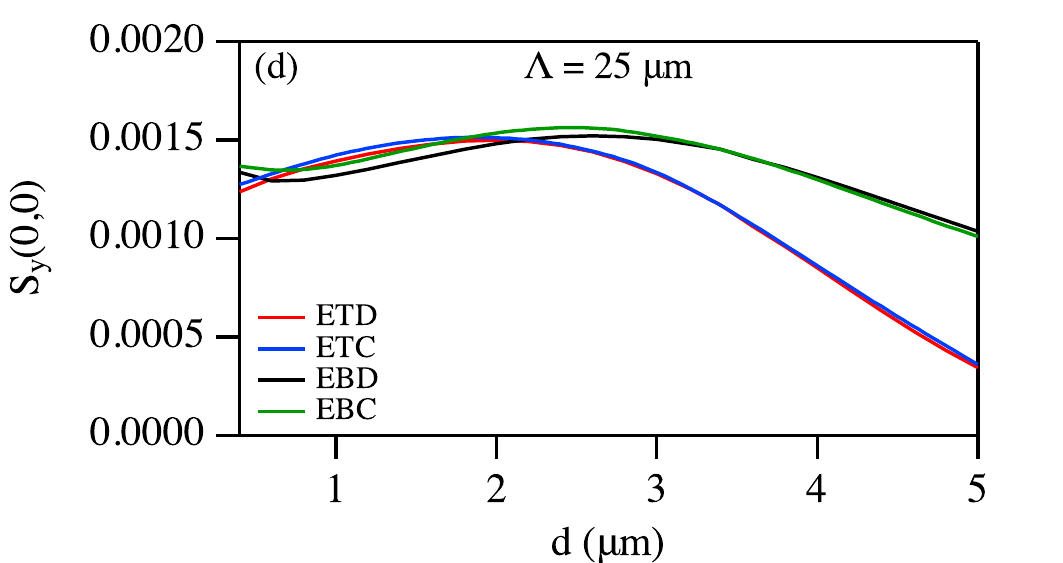}\\
    \end{tabular}
	\caption{Strain in the x-direction (a) at the center of the core, $S_{x}(0,0)$, and corresponding strain in the y-direction (b), $S_{y}(0,0)$ as a function of $\Lambda$ for $d$=1.5, 2.5 and 3.5~$\mu$m. Strain in the y-direction at the center of the core, $S_{y}(0,0)$ versus $\Lambda$ for $d=1.5$~$\mu$m (c) and versus $d$ for $\Lambda = 25$~$\mu$m (d) for the four IDT configurations. A sinusoidal voltage with a resonant frequency and an amplitude of 10~V is applied to the IDT electrode.} 
	\label{fig:S_center}
\end{figure}

In order to quantify how the strain can be maximized via variation of the thickness of the PZT layer and the period of the IDT electrode, we plot in Fig.~\ref{fig:S_center}(a) the induced strain in the $x$-direction at the center of the core, $S_{x}(0,0)$, as a function of the IDT period, $\Lambda$, for $d=$1.5, 2.5 and 3.5~$\mu$m, when a sinusoidal voltage with a resonant frequency and an amplitude of 10 V is applied to the IDT for the configuration \texttt{\footnotesize ETD}. Similarly, Fig.~\ref{fig:S_center}(b) shows the variation of the strain in the $y$-direction at the center of the core, $S_{y}(0,0)$, with the IDT period under the same conditions.	

As the IDT period increases, the acoustic wavelength of the SAW increases as well and the wave penetrates deeper into the structure towards and beyond the core of the optical waveguide. For $S_{y}$ this means that the maximum in its magnitude will move to smaller $y$, \textit{i.e.}, towards larger depth below the PZT film. Therefore, $S_{y}(0,0)$ will first increase, reach a maximum and then decrease as the IDT period, $\Lambda$, increases from 10~$\mu$m upward to 80~$\mu$m, the range investigated. Also the location of zero strain in $x$-direction (see Fig.~\ref{fig:strain}(a)) moves to larger depth (smaller $y$) as the IDT period increases. Consequently, the magnitude of $S_{x}(0,0)$ will first reach a maximum, then it becomes zero and $S_{x}(0,0)$ changes sign for some $\Lambda$, before increasing in magnitude again when the IDT period is further increased.
We observe that only for large periods, typically $\Lambda$ larger than about 50~$\mu$m depending on $d$, do the two strain components have equal sign and cooperate for the strain-optic effect. Unfortunately, at these large periods, the magnitude of both $S_{x}(0,0)$ and $S_{y}(0,0)$ is significantly smaller than the maximum obtained at smaller $\Lambda$ for a given thickness of the PZT film (see Figs.~\ref{fig:S_center}(a) and \ref{fig:S_center}(b)). Therefore, we expect that the strain-optic effect will be significantly reduced at the larger IDT periods, despite the cooperation of the two strain components. Furthermore, a large $\Lambda$ corresponds to a relatively low resonant acoustic frequency (see Fig.~\ref{fig:f_r}). 

Figure~\ref{fig:S_center}(b) indicates that in order to maximize the magnitude of the strain in the $y$-direction at the location of the core for configuration \texttt{\footnotesize ETD} the optimum PZT film thickness needs to increases with increasing IDT period. This is also the case when the magnitude of the strain in the $x$-direction is maximized for the shorter IDT periods ($\Lambda \lesssim 50$~$\mu$m). In order to quantify this further and investigate the effect of the IDT configuration, we plot in Fig.~\ref{fig:S_center}(c) $S_{y}(0,0)$ as a function of $\Lambda$ for $d=1.5$~$\mu$m and in Fig.~\ref{fig:S_center}(d) $S_{y}(0,0)$ as a function of $d$ for $\Lambda = 25$~$\mu$m for the four IDT configurations investigated. 

Figure~\ref{fig:S_center}(c) shows that for $d = 1.5$~$\mu$m all IDT configurations perform approximately equally well for the IDT periods investigated. Whenever the size of the modulator becomes critical, we observe that the configurations with the IDT electrode at the PZT-SiO$_{2}$ interface slightly outperform the configurations with the IDT electrode on top of the PZT film when $\Lambda \lesssim 20$~$\mu$m. On the other hand, when $\Lambda \gtrsim 20$~$\mu$m the configurations with the IDT electrode at the top slightly outperform the ones with the IDT electrode at the PZT-SiO$_{2}$ interface. Further, for $\Lambda \lesssim 20$~$\mu$m there is no significant difference in $S_{y}(0,0)$ when the PZT film is terminated with a conductive film opposite the IDT electrode or not, while for $\Lambda \gtrsim 20$~$\mu$m the configuration with a conductive layer opposite to the IDT electrode slightly outperforms the corresponding one without the conductive layer. 

Keeping the IDT period constant at $\Lambda = 25$~$\mu$m and varying the thickness, $d$, of the PZT film, Fig.~\ref{fig:S_center}(d) shows that all four configurations produce approximately equal strain at the center of the core for $d \lesssim 2.5$~$\mu$m, and that with increasing $d$ beyond this value the configurations with the IDT electrode at the PZT-SiO$_{2}$ interface increasingly outperform the ones with the IDT electrode at the top of the PZT film. Figure~\ref{fig:S_center}(d) also shows that the configurations with a terminating conductive layer opposite to the IDT electrode slightly outperform the corresponding configuration having no conductive layer. It should be noted that as the acoustic wavelength is kept constant, Fig.~\ref{fig:S_center}(d) also reflects the coupling between the voltage on the IDT electrode and strain induced by the SAW.  

In summary, we find that in order to maximize the acoustic modulation frequency, all configurations require a thin PZT layer and a small IDT period. Taking the configuration \texttt{\footnotesize ETD} as reference, terminating the PZT film with a conductive layer opposite to the IDT electrode lowers the resonant frequency, while having the IDT electrode at the PZT-SiO$_{2}$ interface slightly increases the resonance frequency, though generally less than the reduction caused by the conductive layer. However, the need to create maximum strain in the area of the optical mode requires an optimum IDT period with corresponding optimum PZT layer thickness, which will ultimately limit the maximum modulation frequency that can be realized with this configuration. This will be investigated further in the next section.

\subsection{Modulation of the effective refractive index}
As our interest is in a high modulation frequency, we focus on SAWs produced with short period IDTs (\textit{cf.} Fig.~\ref{fig:f_r}). The various configurations considered here produce almost the same strain when $\Lambda$ is varied (see Fig.~\ref{fig:S_center}(c)). Further, configuration \texttt{\footnotesize EBC} produces a strain that in magnitude is about the same or higher compared to that of the other IDT configurations, when $d$ is varied (see Fig.~\ref{fig:S_center}(d)). Therefore, we use configuration \texttt{\footnotesize EBC} to present the relative change in effective refractive index for the fundamental optical mode, which can be TE or TM polarized. 

As shown in the previous section, at modulation frequencies of the order of 100~MHz the SAW-induced strain extends well into the cladding and should be able to cover the whole cross-sectional area occupied by the optical mode. This strain will lead to a change in the refractive index of the cladding and core via Eqs.~(\ref{eq.delta_nx}) and (\ref{eq.delta_ny}), the strength of the coupling being set by the strain-optic coefficients. The strain-optic coefficients are not known for Si$_3$N$_4$, however, due to our choice of a small core area and high aspect ratio, the influence of the strain within the core on the effective refractive index of the optical mode can be neglected. In the model we take the strain-optic coefficients for Si$_3$N$_4$ equal to zero, to obtain a lower bound of the change in effective refractive index that can be realized. For SiO$_2$ we take the strain-optic coefficients to be equal to $p_{11}=0.118$ and $p_{12}=0.252$~\cite{bertholds_determination_1988, heiman_1979,galbraith_2014,schroeder_1980}. Due to the difference in strain in the $x$- and $y$-direction, the refractive index experienced by the mode is different for TE and TM polarization~\cite{fallahkhair_vector_2008}. To find the effective refractive index for the fundamental mode for the two polarization directions, we take the calculated strain and use Eqs.~(\ref{eq.delta_nx}) and (\ref{eq.delta_ny}) to add the appropriate change in refractive index to the material refractive index~\cite{korpel_1996}. Subsequently, we use the eigenmode solver~\cite{comsol_2014} to solve for the effective refractive index $n_{\textrm{eff}}(S)$ for the fundamental  optical mode for both the TE and TM polarization. We then calculate the relative effective refractive index difference $\Delta n/n_{\textrm{eff}} = \left(n_{\textrm{eff}}(S) - n_{\textrm{eff}}\right) /n_{\textrm{eff}}$ where $n_{\textrm{eff}} = n_{\textrm{eff}}(S=0)$ is the effective refractive index for the same mode in absence of the SAW. 

\begin{figure}[t]
	\centering
	\includegraphics[width=0.6\linewidth]{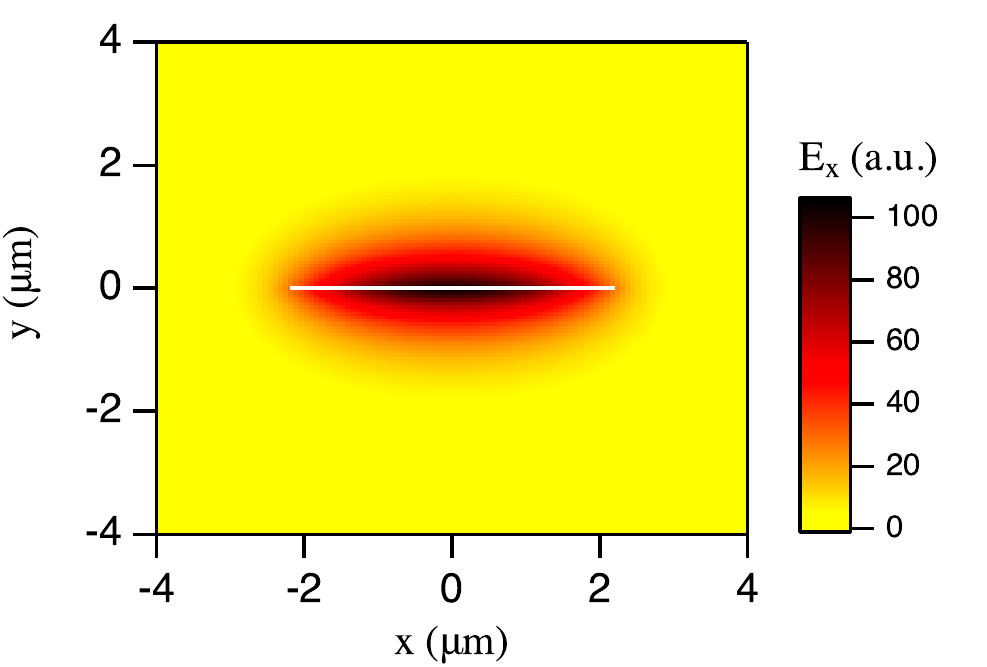}
	\caption{Distribution of the $E_{x}$ component (in arbitrary units) of the fundamental quasi-TE eigenmode for a wavelength of 840~nm. The Si$_3$N$_4$ core with dimensions of 30~nm by 4.4~$\mu$m is centered at the origin of the coordinate system and is indicated by the white line. The drawing is to scale.}
	\label{fig:mode}
\end{figure}

Since the waveguide geometry we choose is meant for visible and near-infrared applications~\cite{haglund_hybrid_2017}, we selected an intermediate wavelength, $\lambda$= 840\-~nm, as an example.
Figure~\ref{fig:mode} shows the $E_{x}$ component of the electric field distribution (in arbitrary units) of the fundamental guided mode with TE polarization using a waveguide core area of $4.4\times 0.03$~$\mu$m$^2$ (shown as white line in the figure) and using the same coordinate system as for Fig.~\ref{fig:strain}. We observe that the mode is confined around the core and already has negligible amplitude for distances a few~$\mu$m away from the core boundary. The effective refractive index for this mode is found to be $n_{\mathrm{\mathrm{eff}}}=1.46630$, which is close to the refractive index of the SiO$_2$ cladding. This confirms that most of the optical field is outside the Si$_3$N$_4$ core and that taking the influence of the strain in the core as negligible is justified. When the SAW-induced strain is applied, the transverse shape of the intensity distribution as displayed in Fig.~\ref{fig:mode} is almost unaffected by the slight change in refractive index of the cladding material, which is of the order of 10$^{-3}$, however, the longitudinal propagation constant is changed and, hence, the effective refractive index.  

\begin{figure}[t]
	\centering
    \begin{tabular} {c c}
	\includegraphics[width=0.45\linewidth]{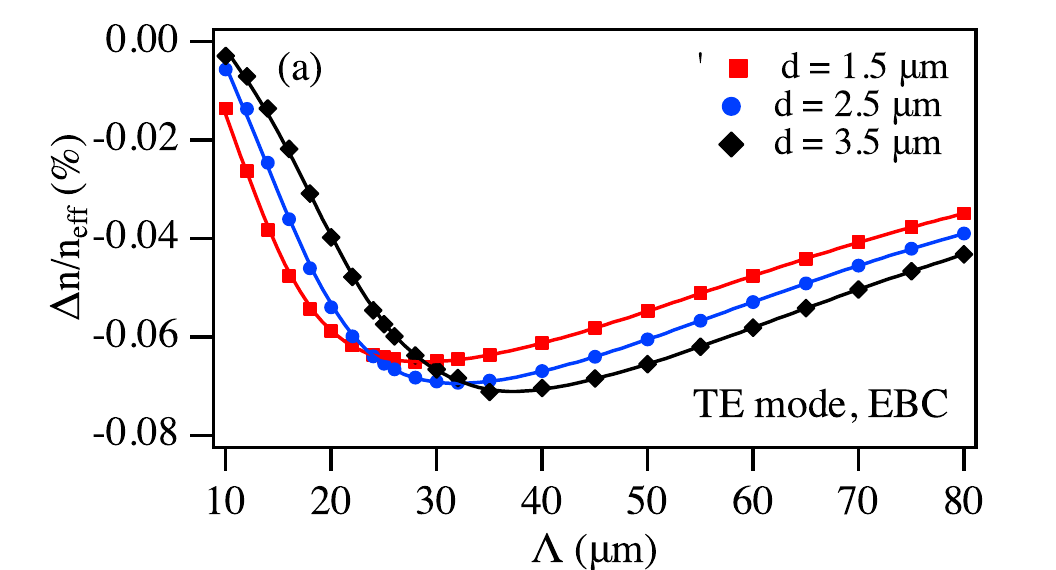} &
    \includegraphics[width=0.45\linewidth]{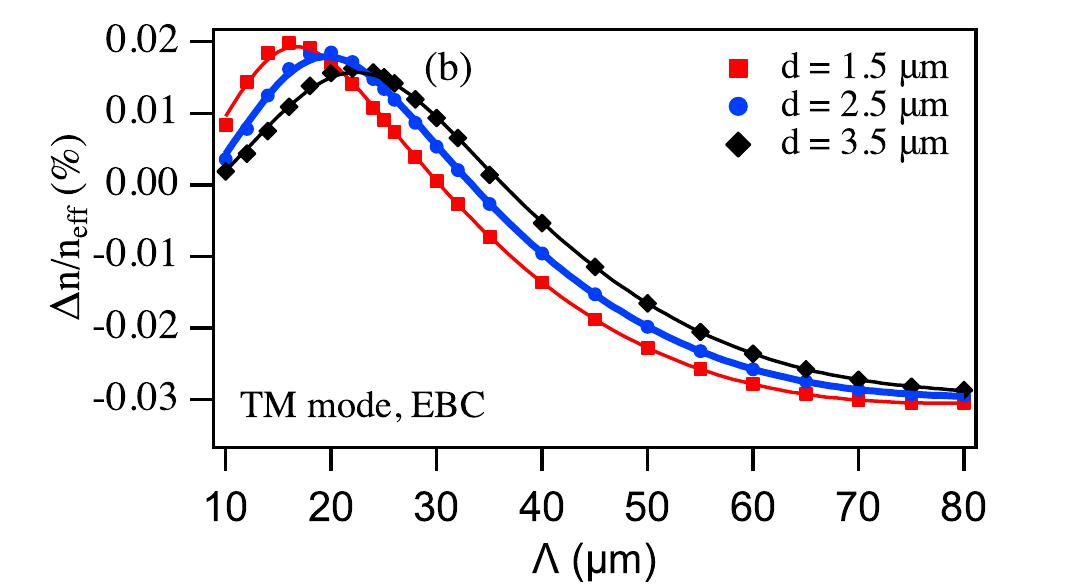} \\
    \includegraphics[width=0.45\linewidth]{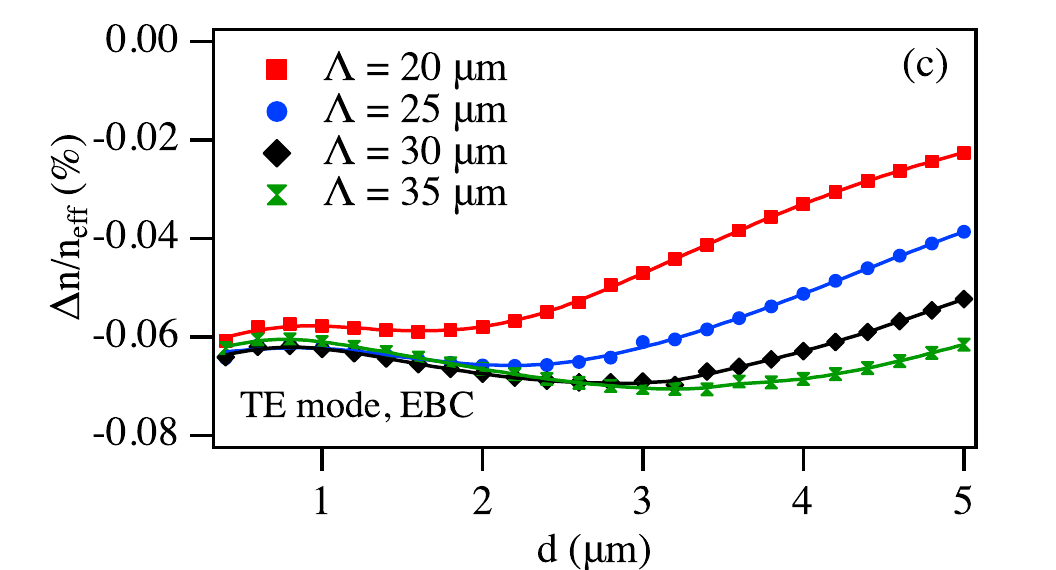} &
    \includegraphics[width=0.45\linewidth]{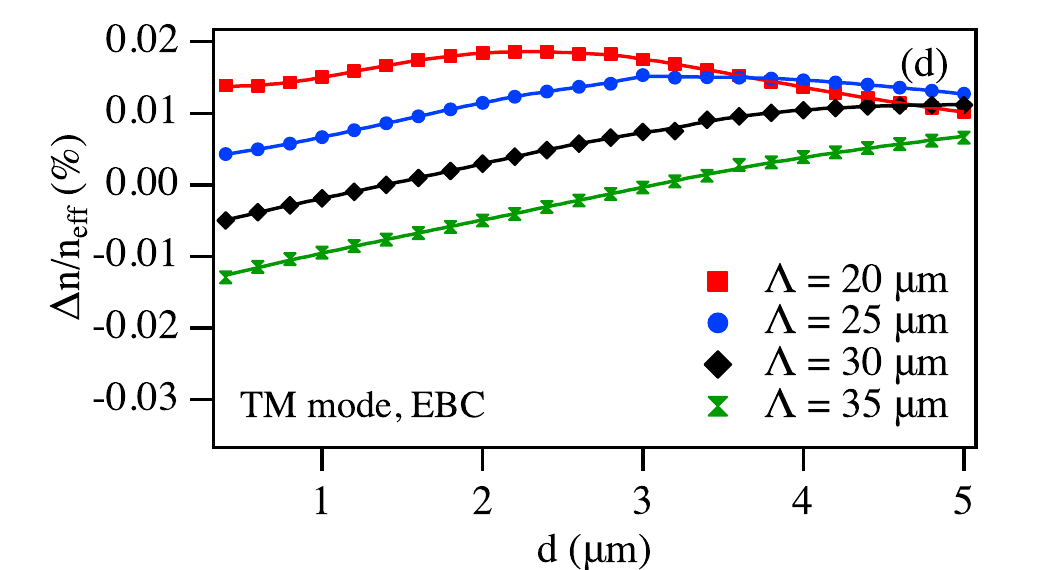} \\
	\end{tabular}
    \caption{Relative change in the effective refractive index, $\Delta n/n_{\mathrm{\mathrm{eff}}}$, for the quasi-TE mode (a) and quasi-TM mode (b) as a function of IDT period, $\Lambda$, for three different PZT-layer thicknesses, $d=1.5$, 2.5 and 3.5~$\mu$m  and as a function of $d$ for $\Lambda=20$, 25, 30 and 35~$\mu$m  for the quasi-TE  mode (c) and for the quasi-TM mode (d). The IDT configuration is \texttt{\footnotesize EBC} and the vacuum wavelength is $\lambda = 840$~nm.}
	\label{fig:dneff}
\end{figure}

\subsubsection*{Variation with IDT period}

In Fig.~\ref{fig:dneff} we show the calculated relative change in effective refractive index, $\Delta n/n_{\mathrm{\mathrm{eff}}}$, for the fundamental mode with TE- (a,c) and TM (b,d) polarization as a function of the IDT period, $\Lambda$, (a,b) and as a function of the PZT film thickness, $d$, (c,d). In (a,b) $\Delta n/n_{\mathrm{\mathrm{eff}}}$ is calculated for three different thicknesses of the PZT layer, $d$ =2.5, 3 and 3.5~$\mu$m, while in (c,d) $\Delta n/n_{\mathrm{\mathrm{eff}}}$ is calculated for four different IDT periods, $\Lambda = 20$, 25, 30 and 35~$\mu$m. In all cases the voltage signal applied to the IDT period is maintained at a constant amplitude of 10~V with the appropriate resonant frequency (see Fig.~\ref{fig:f_r}). 

As the strain-optic coefficient $p_{12}$ is more than a factor of 2 larger than $p_{11}$ for SiO$_{2}$ and because the contribution of the two strain components to the strain-optic effect partly cancel each other at the smaller IDT periods, we expect that TE polarized light will have the dominant shift in $n_{\textrm{eff}}$ in the presence of the SAW. Indeed, Figs.~\ref{fig:dneff}(a) and \ref{fig:dneff}(b) show that the TE polarized fundamental mode has the larger shift in effective refractive index when the SAW is applied. Further, the magnitude of $\Delta n/n_{\mathrm{\mathrm{eff}}}$ for the TE polarized mode reaches its maximum for $\Lambda=35$~$\mu$m and $d=3.5$~$\mu$m, while for the TM polarized mode this is at $\Lambda=80$~$\mu$m or larger for the layer thicknesses investigated. Although the two strain components cooperate in producing the strain-optic effect for the latter case, the magnitude of the relative shift $\Delta n/n_{\mathrm{\mathrm{eff}}}$ is still more than a factor 2 smaller than that obtained with TE polarization.

The magnitude of the two strain components has a local maximum in the range $15$~$\mu$m $\lesssim \Lambda \lesssim 35$~$\mu$m, depending on $d$. Comparing Figs.~\ref{fig:dneff}(a) and \ref{fig:dneff}(b) with Figs.~\ref{fig:S_center}(a) and \ref{fig:S_center}(b) we observe that although this local maximum decreases with increasing thickness of the PZT film, the maximum in the magnitude of the relative shift, $|\Delta n/n_{\mathrm{\mathrm{eff}}}|$, for the TE polarization increases with $d$ (see Fig.~\ref{fig:S_center}(a)). Also the IDT period at which these maxima occur shift towards larger values compared to the values for the local maxima in the strain components. On the other hand, the local maxima in $|\Delta n/n_{\mathrm{\mathrm{eff}}}|$ for the TM polarization decrease with increasing $d$ (see Fig.~\ref{fig:S_center}(b)), like the maxima in the strain. However, for the TM polarization, $|\Delta n/n_{\mathrm{\mathrm{eff}}}|$ reaches its local maximum at IDT period with somewhat smaller values compared to IDT periods for which the magnitude of the strain components is maximum. This behavior is a result of both the spatial overlap of the strain and the optical mode as well as the different weighting of the strain components in the strain-optic effect for the two polarizations. At the smaller IDT periods, the x-component of the strain is dominantly negative in the region of the optical mode (see Fig.~\ref{fig:strain}(a)) and the different weighting by the strain-optic coefficient now causes $\Delta n/n_{\mathrm{\mathrm{eff}}}$ to become positive while it is always negative for the TE polarization. With increasing IDT period, the region of negative $S_{x}$ moves to larger depths and $\Delta n/n_{\mathrm{\mathrm{eff}}}$ first increases due to a stronger coupling of $S_{x}$ with the optical mode. However, when the IDT period further increases, the region of negative $S_{x}$ will start to move out of the region of the optical mode and the region of positive $S_{x}$ will start overlapping with the optical mode. Together with the contribution of $S_{y}$ , this will first reduce and then make $\Delta n/n_{\mathrm{\mathrm{eff}}}$ negative, justs as for the TE polarization. 

Figures~\ref{fig:dneff}(a) and \ref{fig:dneff}(b) show that, so far, the maximum in $|\Delta n/n_{\mathrm{\mathrm{eff}}}|$ is 0.07\% for $d$=3.5~$\mu$m and $\Lambda=35$~$\mu$m, at least for the parameter ranges investigated. In the next section we will consider the variation of $\Delta n/n_{\mathrm{\mathrm{eff}}}$ for the two polarizations with $d$ for a few relevant IDT periods.

\subsubsection*{Variation with thickness of the PZT film}
So far, we have only considered the variation with $\Lambda$ for a few fixed values of the PZT layer thickness. The variation of the calculated $\Delta n/n_{\mathrm{\mathrm{eff}}}$ with thickness, $d$ of the PZT film is shown in Figs.~\ref{fig:dneff}(c) and \ref{fig:dneff}(d) for the TE and TM polarization, respectively. In both cases, $\Delta n/n_{\mathrm{\mathrm{eff}}}$ is calculated for four different IDT periods, $\Lambda = 20$, 25, 30 and 35~$\mu$m for configuration \texttt{\footnotesize EBC} and the remaining parameters are as for Figs.~\ref{fig:dneff}(a) and \ref{fig:dneff}(b). 

Considering first the TE polarization and $\Lambda=20$~$\mu$m, Fig.~\ref{fig:dneff}(c) shows that the change in effective refractive index when the SAW is applied is rather insensitive to $d$ for $d \lesssim 2$~$\mu$m before it starts to drop for values of $d$ beyond this range. On the other hand, for the remaining $\Lambda$ investigated, $|\Delta n/n_{\mathrm{\mathrm{eff}}}|$ first slightly increases before its starts to drop with increasing $d$. The rate at which $|\Delta n/n_{\mathrm{\mathrm{eff}}}|$ drops with increasing $d$ reduces when $\Lambda$ increases. For small $d$ ($d \lesssim 1.5$~$\mu$m) the relative shift $|\Delta n/n_{\mathrm{\mathrm{eff}}}|$ is only weakly dependent on the IDT period. The  two IDT periods $\Lambda = 30$ and 35~$\mu$m obtain both the largest, nearly equal, shift in effective refractive index of $|\Delta n/n_{\mathrm{\mathrm{eff}}}| = 0.07\%$ for $d=3$ and 3.5~$\mu$m, respectively. Furthermore, Fig.~\ref{fig:dneff}(c) also shows that the maxima in $|\Delta n/n_{\mathrm{\mathrm{eff}}}|$ are rather broad, making this device less sensitive to fabrication tolerances in the PZT film thickness. 

The TM polarization shows a different behavior, see Fig.~\ref{fig:dneff}(d). The interplay between $S_{x}$ and $S_{y}$ described above, together with the different weighting by the strain-optic coefficients and the larger penetration depth of the the SAW with increasing period makes that $|\Delta n/n_{\mathrm{\mathrm{eff}}}|$ depends more strongly on the IDT period than in case of TE polarization, for $d \lesssim 1.5$~$\mu$m. The shift in effective refractive index remains smaller for the TM polarization compared to that for the TE polarization for the range of parameters investigated. For a given IDT period, the acoustic wavelength is fixed and the variation with increasing $d$ reflects the coupling between the voltage wave applied to the IDT electrode and the SAW induced strain. For example, for $\Lambda = 25$~$\mu$m, $|\Delta n/n_{\mathrm{\mathrm{eff}}}|$ first increases with $d$ before it starts to drop for $d \gtrsim 3$~$\mu$m, with agrees with the variation of $S_{y}(0,0)$ with $d$, see Fig.~\ref{fig:S_center}(d). 

In summary, We find that the largest $|\Delta n/n_{\mathrm{\mathrm{eff}}}|$ of $\sim0.07$\% is obtained with IDT configuration \texttt{\footnotesize EBC} for $\Lambda\approx30$ - 35~$\mu$m and $d\approx2.5$ - 3.5~$\mu$m. This corresponds to an absolute change in index of $\Delta$n = 1.0~$\times$~10$^{-3}$. The other IDT configurations produce nearly the same shift in the effective refractive index, except when $d$ becomes large. However, a large $d$ requires a larger $\Lambda$ for optimum generation of the SAW and this means that the maxima in the magnitude of the strain will be located below the core of the optical waveguide, resulting in a reduced strain at the location of the optical mode. Also, the acoustic, and therefore also the modulation, frequency will be low. Hence, we will limit ourselves to $d \lesssim 3$~$\mu$m. Inducing a maximum shift therefore requires choosing (i) an appropriate period of the IDT and (ii) the corresponding optimum thickness of the PZT layer. To find the maximum value for $\Delta n/n_{\mathrm{\mathrm{eff}}}$ requires a two-dimensional scan over the IDT period and PZT layer thickness for each of the configurations. Although we have not fully scanned the complete parameter space, Fig.~\ref{fig:dneff} indicates that the scans presented in this figure should be close to or even contain the optimum combination of $\Lambda$ and $d$ to achieve a maximum change in the effective refractive index.

\subsubsection*{Variation with thickness top cladding}
So far, we have considered a standard configuration for the optical waveguide consisting of a Si$_{3}$N$_{4}$ core in the middle of a 16~$\mu$m thick SiO$_{2}$ cladding. Figure~\ref{fig:mode} suggests that the optical field is confined to an area close to the core, at least for the 840~nm wavelength we are considering here. Therefore, it should be possible to reduce the cladding height above the core to improve the coupling between SAW induced strain and the optical mode without increasing the propagation losses of the optical mode. This is especially of interest for the shorter IDT periods, where the penetration of the SAW into the cladding is limited due to the small acoustic wavelength. Further, we found that for the shorter IDT periods $S_{x}$ counteracts $S_{y}$ in the strain-optic effect, \textit{e.g.}, see Fig.~\ref{fig:strain}, and moving the core closer to the PZT film would both reduce $|S_{x}|$ and increase $|S_{y}|$. We therefore expect a stronger strain-optic effect. 

\begin{figure}[b]
	\centering
	\includegraphics[width= 0.6 \linewidth]{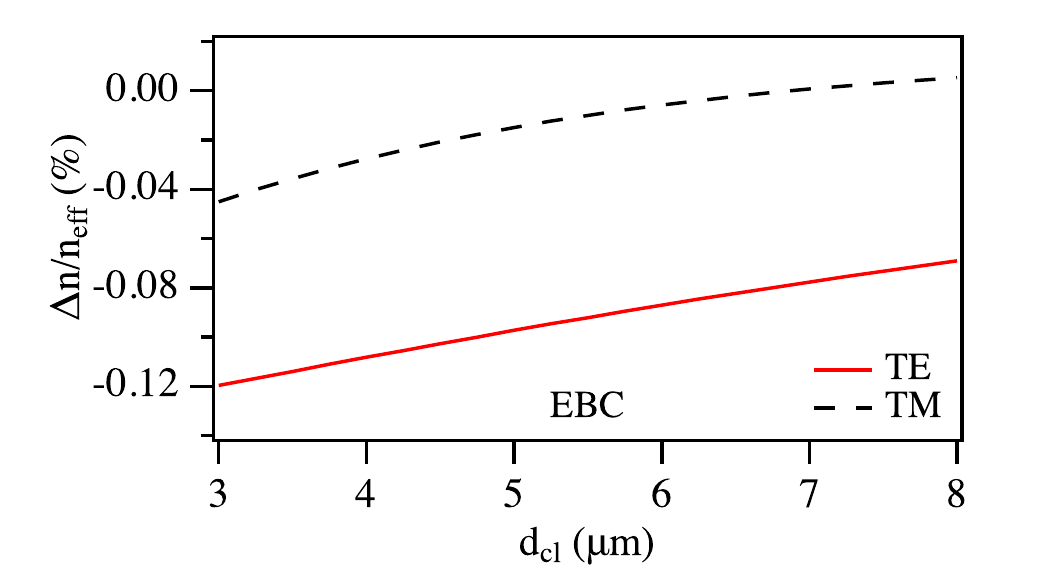}
	\caption{\label{fig:neff_dcladding}
		Relative change in the effective refractive index, $\Delta n/n_{\mathrm{\mathrm{eff}}}$, as a function thickness of the cladding layer above the core, $d_{\mathrm{cl}}$  for the fundamental mode with TE polarization (red line) and TM polarization (dashed black line). The IDT configuration is \texttt{\footnotesize EBC}, $d=2.5$~$\mu$m, $\Lambda=30$~$\mu$m and a voltage signal with an amplitude of 10~V and the appropriate resonant frequency is applied to the IDT electrodes. The vacuum wavelength is $\lambda = 840$~nm.
	}
\end{figure}

In Fig.~\ref{fig:neff_dcladding} we plot the relative shift $\Delta n/n_{\mathrm{\mathrm{eff}}}$ for both the TE polarization (solid red line) and TM polarization (dashed black line) as a function of the thickness of the cladding above the core, $d_{\textrm{cl}}$ . We consider again IDT configuration \texttt{\footnotesize EBC}. The dimensions of the core are the same as in Fig.~\ref{fig:dneff}, the vacuum wavelength is again 840~nm and $d=2.5$~$\mu$m and $\Lambda=30$~$\mu$m. At the smallest height of $d_{\textrm{cl}}$ of 3~$\mu$m investigated and with no SAW applied, we calculate that the effective refractive index of the fundamental mode differs by a small fraction of $2\times 10^{-6}$ from the effective refractive index for the standard configuration used for Fig.~\ref{fig:dneff}, which is about three orders of magnitude smaller than the largest shift induced by the SAW wave. Therefore, we conclude that nearness of the boundary (the PZT-SiO$_{2}$ interface) had little effect on the eigenmode found by the modal solver and that the optical mode is still well confined to the cladding and no substantial increase in propagation loss is expected. We have verified this by calculating the propagation loss for the core of Fig.~\ref{fig:mode}, symmetrically embedded in a cladding of full height $2d_{\textrm{cl}}$ and terminated with the Si substrate on one side and air on the other side. We find that, at the wavelength considered, the propagation loss is less than 0.1~dB/cm for $d_{\textrm{cl}}\ge3.5$~$\mu$m and increases to 10.8~dB/cm when $d_{\textrm{cl}}$ is reduced to 2~$\mu$m, slightly more than doubling for every decrease of 0.25~$\mu$m in $d_{\textrm{cl}}$.
 
Figure~\ref{fig:neff_dcladding} shows that the relative shift in effective refractive index almost doubles from -0.07\% to -0.12\% for the TE polarization when $d_{\textrm{cl}}$ is reduced from 8~$\mu$m to 3~$\mu$m. For the TM polarization, the initial positive shift, caused by the dominant contribution of $S_{x}$ in the strain-optic effect, changes to a negative shift of slightly more than -0.04\% when $d_{\textrm{cl}}$ is varied in the same way. Moving the core closer to the PZT film reduces the contribution from $S_{x}$ to the strain-optic effect (see Fig.~\ref{fig:strain}), until at around $d_{\textrm{cl}}=3$~$\mu$m, it is mainly $S_{y}$ that produces the strain-optic effect. The difference in weighting by the strain-optic coefficients produces the smaller SAW-induced shift for the TM polarization (\textit{c.f.} Eqs.~(\ref{eq.delta_nx}) and (\ref{eq.delta_ny})). 

\subsubsection*{Optimized design and geometry}
Now that we have calculated the maximum value for $|\Delta n/n_{\mathrm{\mathrm{eff}}}|$ provided by the optimal configuration, which is 0.07\% for the standard configuration ($d_{\textrm{cl}}=8$~$\mu$m) and 0.12\% for the configuration with $d_{\textrm{cl}}=3$~$\mu$m, we use Eq.~(\ref{eq.length}) to calculate the required length of the arms of a balanced Mach-Zehnder interferometer to obtain complete light modulation at the acoustic resonance frequency. For a vacuum wavelength of $\lambda=840$~nm, we find $L=205\,(120)$~$\mu$m using $n_{\mathrm{\mathrm{eff}}}=1.46630$ for TE polarized light and $d_{\textrm{cl}}=8\,(3)$~$\mu$m.  The corresponding figure-of-merit, $V_{\pi}L$, is 0.4\,(0.24)~Vcm. The voltage applied to the IDT has an amplitude of 10~V. Due to the linearity of the strain with applied voltage expected in this regime of small strain and index modulation, obtaining full light modulation at a shorter wavelength would require driving the IDT at a smaller amplitude. In contrast, modulating light with a longer wavelength, say at telecommunication wavelengths (1550~nm), would almost double the required driving voltage ($\sim 18.5$~V). Nevertheless, both values are well within the expected operating range that may extend to voltage amplitudes of 50~V or more before breakdown occurs~\cite{deLima_2006, hashimoto_2000}. This means that depending on the selected voltage or wavelength even a smaller arm length than 205~(120)~$\mu$m might be sufficient to obtain full light modulation. Also, balancing refractive index modulation against acceptable propagation loss may further reduce the device length or relax the driving voltage requirements. 

\section{Summary and conclusions}
In this work we investigated the use of Rayleigh-type surface acoustic waves (SAWs) to modulate the effective refractive index of an optical mode guided by a buried Si$_3$N$_4$ waveguide core in a SiO$_2$ cladding. We considered that the acoustic waves are excited in a PZT piezo-electric layer deposited on top of the waveguide cladding via interdigitized electrodes, at a frequency of the order of 100~MHz. 

Considering a balanced Mach-Zehnder interferometer, the modulation of the effective refractive index can be used to obtain full, \textit{i.e.}, 100\%-modulation of the light power and amplitude, at the acoustic frequency. The optical waveguide considered here consists of a Si$_3$N$_4$ core, with dimensions of 4.4~$\mu$m by 30~nm, buried in a SiO$_2$ cladding $8$~$\mu$m below the surface, which is typical for this low-loss photonic platform. The SAWs generated by the thin PZT layer is guided in the interface between the PZT and the cladding, while its evanescent strain field extends towards depths that include the waveguide core. The strain induced by the SAW results in a change of the effective refractive index of the waveguide via the strain-optic effect. 

We find that of four investigated IDT-PZT arrangements, the configuration with the IDT electrode at the PZT-SiO$_2$ interface and having a terminating conducting film  at the PZT-air interface (configuration \texttt{\footnotesize EBC}) is the most efficient in generating strain in the cross sectional area of the optical mode, although the difference with the other IDT geometries is small when operating at high modulation frequencies. The induced strain produces the largest strain-optic effect for TE polarized light. For this polarization, a maximum relative change in effective refractive index for the fundamental waveguide mode of $|\Delta n/n_{\mathrm{eff}}|=0.07$\%  is found for a wavelength in the middle of the working range of this waveguide, here taken as $\lambda=840$~nm. The strain-optic effect increases to $|\Delta n/n_{\mathrm{eff}}|=0.12$\% when the height of the cladding above the core is reduced to 3~$\mu$m. These values represent a lower bound as we have set the strain-optic coefficients to zero inside the core. However, as most of the optical field is outside the core, the difference with the actual value should be small.  For the standard waveguide configuration ($d_{\textrm{cl}}=8$~$\mu$m) and using a driving voltage amplitude of 10~V, the maximum modulation is obtained at resonance for an IDT  period of  $\Lambda \approx30$ - 35~$\mu$m and a PZT film thickness $d \approx 2.5$ - 3.5~$\mu$m, while the resonant frequency is in the range 70-85~MHz. The dependence on $d_{\mathrm{cl}}$ is only investigated for $d=2.5$~$\mu$m and $\Lambda = 30$~$\mu$m, and the resonant frequency is 87.3~MHz for the optimum $d_{\textrm{cl}}$ of 3~$\mu$m.

For the maximum relative change in refractive index, the arm length required in a balanced Mach-Zehnder interferometer is $L=205\,(120)$~$\mu$m , for 100\% light modulation using a voltage signal with a 10-V amplitude at a frequency of 84.5~(87.3)~MHz, when the height of the cladding above the core is $d_{\textrm{cl}}=8\,(3)$~$\mu$m, $d=2.5$~$\mu$m and $\Lambda = 30$~$\mu$m. We note that this frequency is larger by at about five orders of magnitude compared to thermo-optic intensity modulators and about two orders of magnitude compared to modulators based on non-resonant proximity piezo-actuation~\cite{hosseini_2015}. The figure-of-merit, $V_{\pi}L$, equals to 0.4~Vcm and 0.24~Vcm for $d_{\textrm{cl}} = 8$ and 3~$\mu$m, respectively.  

We note that also the required interaction length (for a MZI) is shorter, by a factor about five to ten, than what is typically used in thermally operated MZI (500~$\mu$m) and by a factor of about 100 compared to proximity strain-optic modulation~\cite{hosseini_2015}. The figure-of-merit, $V_{\pi}L$, for the SAW modulator is on par with electro-optic modulators using a BaTiO$_3$ thin film~\cite{tang_2005}, however it is much better, \textit{i.e.}, lower, than the figure-of-merit for PZT-on-Si$_3$N$_4$ based ring-modulators~\cite{alexander_2018} when corrected for the different wavelengths used. Although the design presented here does not reach the very high modulation frequencies (tens of GHz) obtained by the electro-optic modulators, this SAW based modulator should have much lower propagation loss, as the optical field remains confined to the SiO$_2$ cladding and the Si$_3$N$_4$ core.  

As a parallel route for optimization, IDT electrodes might be configured to generate a focused SAW~\cite{collins_2016} to increase the strain in the region of the optical mode. An additional variation would be meandering the optical waveguide though the SAW field for shortening of the required overlap length with the transverse SAW field dimension.Also, balancing refractive index modulation against acceptable propagation loss may further improve, \textit{i.e.}, lower $V_{\pi}L$. 
Another method to decrease to distance between the core and PZT film would be to increase the volume (\textit{i.e.}, the thickness) of the core as this confines the optical mode more to the core and allows better positioning of the core and increases the strain-optic effect. Furthermore, the propagating nature of SAWs can be exploited to modulate light in a waveguide that is located at some distance from a PZT strip with an IDT.  This allows to optimize the waveguide core location with respect to the SAW-induced strain without increasing the propagation loss. As only a few acoustic wavelengths are required to separate the optical field from the PZT-IDT structure, little damping of the acoustic wave is expected (the acoustic eigenmode calculation gives an acoustic power damping of approximately 1 dB/cm for $\Lambda=30$~$\mu$m, $d=1.5$~$\mu$m and using the damping given in Table~\ref{tab:materials}). This is especially advantageous when optimizing the system for high modulation frequency.
Another advantage of using a SAW to drive a Mach-Zehnder interferometer is that it can coherently drive multiple interferometers located suitably next to each other for providing a stable phasing relative to each other. This is of interest, \textit{e.g.}, for low-loss phase modulators that form optical isolators based on acoustic waves~\cite{hendrickson_integrated_2014, shi_acousto-optic_2017}. 
	 	
\section*{Acknowledgment}
The authors would like to thank LioniX International B.V., Enschede, The Netherlands for providing the material data required for calculation of the optical propagation loss.
	 	
\section*{Funding}
This research is supported by NanoNextNL (6B-Functional Nanophotonics), a micro- and nanotechnology consortium of the Government of the Netherlands and 130 partners, and the Netherlands Organization for Scientific Research, NWO, (STW project 11358), which is partly funded by the Ministry of Economic Affairs.



\begin{thebibliography}{10}
	\newcommand{\enquote}[1]{``#1''}
	
	\bibitem{bauters_ultra-low-loss_2011}
	J.~F. Bauters, M.~J.~R. Heck, D.~John, D.~Dai, M.-C. Tien, J.~S. Barton,
	A.~Leinse, R.~G. Heideman, D.~J. Blumenthal, and J.~E. Bowers,
	\enquote{Ultra-low-loss high-aspect-ratio {Si}$_{3}${N}$_{4}$ waveguides,}
	Opt. Express \textbf{19}, 3163--3174 (2011).
	
	\bibitem{worhoff_2015}
	K.~W\"{o}rhoff, R.~Heideman, A.~Leinse, and M.~Hoekman, \enquote{Triplex: A
		versatile dielectric photonic platform,} Adv. Opt. Techn. \textbf{4},
	189--207 (2015).
	
	\bibitem{heck_2014}
	M.~J.~R. Heck, J.~F. Bauters, M.~L. Davenport, D.~T. Spencer, and J.~E. Bowers,
	\enquote{Ultra-low loss waveguide platform and its integration with silicon
		photonics,} Laser Photon. Rev. \textbf{8}, 667--686 (2014).
	
	\bibitem{taballione_2018}
	C.~Taballione, T.~A.~W. Wolterink, J.~Lugani, A.~Eckstein, B.~A. Bell,
	R.~Grootjans, I.~Visscher, D.~Geskus, C.~G.~H. Roeloffzen, J.~J. Renema,
	I.~A. Walmsley, P.~W.~H. Pinkse, and K.-J. Boller, \enquote{8x8 programmable
		quantum photonic processor based on silicon nitride waveguides,} arxiv.org 
	1805.10999 (2018).
	
	\bibitem{epping_integrated_2013}
	J.~P. Epping, M.~Kues, P.~J.~M. van~der Slot, C.~J. Lee, C.~Fallnich, and K.-J.
	Boller, \enquote{Integrated {CARS} source based on seeded four-wave mixing in
		silicon nitride,} Opt. Express \textbf{21}, 32123--32129 (2013).
	
	\bibitem{nguyen_2012}
	V.~D. Nguyen, N.~Weiss, W.~Beeker, M.~Hoekman, A.~Leinse, R.~G. Heideman, T.~G.
	van Leeuwen, and J.~Kalkman, \enquote{Integrated-optics-based swept-source
		optical coherence tomography,} Opt. Lett. \textbf{37}, 4820--4822 (2012).
	
	\bibitem{zinoviev_2008}
	K.~Zinoviev, L.~G. Carrascosa, J.~S\'{a}nchez~del R\'{i}o, B.~Sep\'{u}lveda,
	C.~Dom\'{i}nguez, and L.~M. Lechuga, \enquote{Silicon {Photonic} {Biosensors}
		for {Lab}-on-a-{Chip} {Applications},} Adv. Opt. Technol. \textbf{2008},
	383927 (2008).
	
	\bibitem{roeloffzen_2013}
	C.~G.~H. Roeloffzen, L.~Zhuang, C.~Taddei, A.~Leinse, R.~G. Heideman, P.~W.~L.
	van Dijk, R.~M. Oldenbeuving, D.~A.~I. Marpaung, M.~Burla, and K.-J. Boller,
	\enquote{Silicon nitride microwave photonic circuits,} Opt. Express
	\textbf{21}, 22937--22961 (2013).
	
	\bibitem{marpaung_integrated_2013}
	D.~Marpaung, C.~Roeloffzen, R.~Heideman, A.~Leinse, S.~Sales, and J.~Capmany,
	\enquote{Integrated microwave photonics,} Laser Photon. Rev. \textbf{7},
	506--538 (2013).
	
	\bibitem{oldenbeuving_2013}
	R.~M. Oldenbeuving, E.~J. Klein, H.~L. Offerhaus, C.~J. Lee, H.~Song, and K.-J.
	Boller, \enquote{25 khz narrow spectral bandwidth of a wavelength tunable
		diode laser with a short waveguide-based external cavity,} Laser Phys. Lett.
	\textbf{10}, 015804 (2013).
	
	\bibitem{fan_2017}
	Y.~Fan, R.~M. Oldenbeuving, C.~G. Roeloffzen, M.~Hoekman, D.~Geskus, R.~G.
	Heideman, and K.-J. Boller, \enquote{290 {Hz} intrinsic linewidth from an
		integrated optical chip-based widely tunable {InP-Si}$_{3}${N}$_{4}$ hybrid
		laser,} in \enquote{Conference on Lasers and Electro-Optics,}  (Optical
	Society of America, 2017), JTh5C.9.
	
	\bibitem{halir_ultrabroadband_2012}
	R.~Halir, Y.~Okawachi, J.~S. Levy, M.~A. Foster, M.~Lipson, and A.~L. Gaeta,
	\enquote{Ultrabroadband supercontinuum generation in a {CMOS}-compatible
		platform,} Opt. Lett. \textbf{37}, 1685--1687 (2012).
	
	\bibitem{epping_-chip_2015}
	J.~P. Epping, T.~Hellwig, M.~Hoekman, R.~Mateman, A.~Leinse, R.~G. Heideman,
	A.~van Rees, P.~J.~M. van~der Slot, C.~J. Lee, C.~Fallnich, and K.-J. Boller,
	\enquote{On-chip visible-to-infrared supercontinuum generation with more than
		495 {THz} spectral bandwidth,} Opt. Express \textbf{23}, 19596--19604 (2015).
	
	\bibitem{zhuang_2014}
	L.~Zhuang, M.~Hoekman, C.~Taddei, A.~Leinse, R.~G. Heideman, A.~Hulzinga,
	J.~Verpoorte, R.~M. Oldenbeuving, P.~W.~L. van Dijk, K.-J. Boller, and
	C.~G.~H. Roeloffzen, \enquote{On-chip microwave photonic beamformer circuits
		operating with phase modulation and direct detection,} Opt. Express
	\textbf{22}, 17079--17091 (2014).
	
	\bibitem{xiong_2015}
	C.~Xiong, X.~Zhang, A.~Mahendra, J.~He, D.-Y. Choi, C.~J. Chae, D.~Marpaung,
	A.~Leinse, R.~G. Heideman, M.~Hoekman, C.~G.~H. Roeloffzen, R.~M.
	Oldenbeuving, P.~W.~L. van Dijk, C.~Taddei, P.~H.~W. Leong, and B.~J.
	Eggleton, \enquote{Compact and reconfigurable silicon nitride time-bin
		entanglement circuit,} Optica \textbf{2}, 724--727 (2015).
	
	\bibitem{zhuang_2015}
	L.~Zhuang, C.~G.~H. Roeloffzen, M.~Hoekman, K.-J. Boller, and A.~J. Lowery,
	\enquote{Programmable photonic signal processor chip for radiofrequency
		applications,} Optica \textbf{2}, 854--859 (2015).
	
	\bibitem{ovvyan_2016}
	A.~Ovvyan, N.~Gruhler, S.~Ferrari, and W.~Pernice, \enquote{Cascaded
		{Mach}-{Zehnder} interferometer tunable filters,} J. Opt. \textbf{18}, 064011
	(2016).
	
	\bibitem{tang_2005}
	P.~Tang, A.~L. Meier, D.~J. Towner, and B.~W. Wessels, \enquote{Batio$_{3}$
		thin-film waveguide modulator with a low voltage--length product at
		near-infrared wavelengths of 0.98 and 1.55 {\textmu}m,} Opt. Lett.
	\textbf{30}, 254--256 (2005).
	
	\bibitem{tang_2005b}
	P.~Tang, A.~L. Meier, D.~J. Towner, and B.~W. Wessels, \enquote{High-speed
		travelling-wave batio$_3$ thin-film electro-optic modulators,} Electron.
	Lett. \textbf{41}, 1296--1297 (2005).
	
	\bibitem{alexander_2018}
	K.~Alexander, J.~P. George, J.~Verbist, K.~Neyts, B.~Kuyken, D.~V. Thourhout,
	and J.~Beeckman, \enquote{Nanophotonic pockels modulators on a silicon
		nitride platform,} Nat. Commun. \textbf{9}, 3444 (2018).
	
	\bibitem{tsai_guided-wave_2013}
	C.~S. Tsai, ed., \emph{Guided-wave acousto-optics: interactions, devices, and
		applications} (Springer Verlag, 2013).
	
	\bibitem{chang_2017}
	L.~Chang, M.~H.~P. Pfeiffer, N.~Volet, M.~Zervas, J.~D. Peters, C.~L.
	Manganelli, E.~J. Stanton, Y.~Li, T.~J. Kippenberg, and J.~E. Bowers,
	\enquote{Heterogeneous integration of lithium niobate and silicon nitride
		waveguides for wafer-scale photonic integrated circuits on silicon,} Opt.
	Lett. \textbf{42}, 803--806 (2017).
	
	\bibitem{brewster_1815}
	D.~Brewster, \enquote{On the effects of simple pressure in producing that
		species of crystallization which forms two oppositely polarised images, and
		exhibits the complementary colours by polarised light,} Phil. Trans. R. Soc.
	Lond. \textbf{105}, 60--64 (1815).
	
	\bibitem{yariv_2007}
	A.~Yariv and P.~Yeh, \emph{Photonics: Optical Electronics in Modern
		Communications} (Oxford University, 2007).
	
	\bibitem{hosseini_2015}
	N.~Hosseini, R.~Dekker, M.~Hoekman, M.~Dekkers, J.~Bos, A.~Leinse, and
	R.~Heideman, \enquote{Stress-optic modulator in triplex platform using a
		piezoelectric lead zirconate titanate (pzt) thin film,} Opt. Express
	\textbf{23}, 14018--14026 (2015).
	
	\bibitem{epping_2017}
	J.~P. Epping, D.~Marchenko, A.~Leinse, R.~Mateman, M.~Hoekman, L.~Wevers, E.~J.
	Klein, C.~G.~H. Roeloffzen, M.~Dekkers, and R.~G. Heideman,
	\enquote{Ultra-low-power stress-optics modulator for microwave photonics,}
	Proc.SPIE \textbf{10106}, {101060F} (2017).
	
	\bibitem{schriever_2012}
	C.~Schriever, C.~Bohley, J.~Schilling, and R.~B. Wehrspohn, \enquote{{Strained
			Silicon Photonics},} {Materials} \textbf{{5}}, {889--908} (2012).
	
	\bibitem{deLima_2006}
	M.~M. de~Lima, Jr., M.~Beck, R.~Hey, and P.~V. Santos, \enquote{{Compact
			Mach-Zehnder acousto-optic modulator},} {Appl. Phys. Lett.} \textbf{{89}},
	121104 (2006).
	
	\bibitem{tadesse_2014}
	S.~A. Tadesse and M.~Li, \enquote{Sub-optical wavelength acoustic wave
		modulation of integrated photonic resonators at microwave frequencies,} Nat.
	Commun. \textbf{5}, 5402 (2014).
	
	\bibitem{tadesse_2015}
	S.~A. Tadesse, H.~Li, Q.~Liu, and M.~Li, \enquote{Acousto-optic modulation of a
		photonic crystal nanocavity with lamb waves in microwave k band,} Appl. Phys.
	Lett. \textbf{107}, 201113 (2015).
	
	\bibitem{rayleigh_waves_1885}
	L.~Rayleigh, \enquote{On waves propagated along the plane surface of an elastic
		solid,} Proc. Lond. Math. Soc. \textbf{s1-17}, 4--11 (1885).
	
	\bibitem{lima_modulation_2005}
	M.~M.~d. Lima and P.~V. Santos, \enquote{Modulation of photonic structures by
		surface acoustic waves,} Rep. Prog. Phys. \textbf{68}, 1639--1701 (2005).
	
	\bibitem{kay_1955}
	H.~Kay, \enquote{Electrostriction,} Rep. Prog. Phys. \textbf{18}, 230--250
	(1955).
	
	\bibitem{sundar_1992}
	V.~Sundar and R.~E. Newnham, \enquote{Electrostriction and polarization,}
	Ferroelectrics \textbf{135}, 431--446 (1992).
	
	\bibitem{boyd_nonlinear_2013}
	R.~W. Boyd, \emph{Nonlinear Optics} (Academic Press, 2013).
	
	\bibitem{borrelli_1968}
	N.~F. Borrelli and R.~A. Miller, \enquote{Determination of the individual
		strain-optic coefficients of glass by an ultrasonic technique,} Appl. Opt.
	\textbf{7}, 745--750 (1968).
	
	\bibitem{korpel_1996}
	A.~Korpel, \emph{Acousto-{Optics}} (CRC Press, 1996).
	
	\bibitem{rebiere_1992}
	D.~Rebi\'{e}re, J.~Pistr\`{e}, M.~Hoummady, D.~Hauden, P.~Cunin, and
	R.~Planade, \enquote{Sensitivity comparison between gas sensors using saw and
		shear horizontal plate-mode oscillators,} Sens. Actuator B-Chem. \textbf{6}, 274--278 (1992).
	
	\bibitem{arnau_2008}
	A.~Arnau, ed., \emph{Piezoelectric Transducers and Applications}
	(Springer-Verlag, 2008).
	
	\bibitem{telford_1990}
	W.~Telford, L.~Geldart, and R.~Sheriff, \emph{Applied Geophysics} (Cambridge
	University, 1990).
	
	\bibitem{hashimoto_2000}
	K.~Hashimoto, \emph{Surface aucoustic devices in telecommunications: modelling
		and simulation} (Springer-Verlag, 2000).
	
	\bibitem{saito_2004}
	Y.~Saito, H.~Takao, T.~Tani, T.~Nonoyama, K.~Takatori, T.~Homma, T.~Nagaya, and
	M.~Nakamura, \enquote{Lead-free piezoceramics,} Nature \textbf{432}, 84--87
	(2004).
	
	\bibitem{fukushima_1984}
	J.~Fukushima, K.~Kodaira, and T.~Matsushita, \enquote{Preparation of
		ferroelectric pzt films by thermal decomposition of organometallic
		compounds,} J. Mater. Sci. \textbf{19}, 595--598 (1984).
	
	\bibitem{comsol_2014}
	\url{www.comsol.com}, \enquote{{COMSOL} {M}ultiphysics {V}5.0}   (2014).
	
	\bibitem{luke_broadband_2015}
	K.~Luke, Y.~Okawachi, M.~R.~E. Lamont, A.~L. Gaeta, and M.~Lipson,
	\enquote{Broadband mid-infrared frequency comb generation in a
		{Si}$_{3}${N}$_{4}$ microresonator,} Opt. Lett. \textbf{40}, 4823--4826
	(2015).
	
	\bibitem{lazan_damping_1968}
	B.~J. Lazan, \emph{Damping of materials and members in structural mechanics}
	(Pergamon Press, 1968).
	
	\bibitem{zhang_2014}
	L.~Zhang, R.~Barrett, P.~Cloetens, C.~Detlefs, and M.~Sanchez~del Rioa,
	\enquote{Anisotropic elasticity of silicon and its application to the
		modelling of x-ray optics,} J. Synchrotron Radiat. \textbf{21}, 507--517
	(2014).
	
	\bibitem{morgan}
	D.~Berlincourt, H.~H.~A. Krueger, and C.~Near, \enquote{Properties of
		piezoelectricity ceramics,} Morgan Electro Ceramics, technical
	publication TP-226.
	
	\bibitem{bertholds_determination_1988}
	A.~Bertholds and R.~Dandliker, \enquote{Determination of the individual
		strain-optic coefficients in single-mode optical fibres,} J. Light. Technol.
	\textbf{6}, 17--20 (1988).
	
	\bibitem{heiman_1979}
	D.~Heiman, D.~Hamilton, and R.~Hellwarth, \enquote{Brillouin scattering
		measurements on optical glasses,} Phys. Rev. B \textbf{19}, 6583 -- 6592
	(1979).
	
	\bibitem{galbraith_2014}
	J.~Galbraith, \emph{Photoelastic properties of oxide and non-oxide glasses}
	Ph.D. thesis, Dalhousic University, Nova Scotia (2014).
	
	\bibitem{schroeder_1980}
	J.~Schroeder, \enquote{Brillouin scattering and pockels coefficients in
		silicate glasses,} J. Non-Cryst. Solids \textbf{40}, 549 --566 (1980).
	
	\bibitem{fallahkhair_vector_2008}
	A.~B. Fallahkhair, K.~S. Li, and T.~E. Murphy, \enquote{Vector finite
		difference modesolver for anisotropic dielectric waveguides,} J. Light.
	Technol. \textbf{26}, 1423--1431 (2008).
	
	\bibitem{haglund_hybrid_2017}
	E.~P. Haglund, S.~Kumari, J.~S. Gustavsson, E.~Haglund, G.~Roelkens, R.~G.
	Baets, and A.~Larsson, \enquote{Hybrid vertical-cavity laser integration on
		silicon,} Proc. SPIE \textbf{10122}, 101220H (2017).
	
	\bibitem{collins_2016}
	D.~J. Collins, A.~Neild, and Y.~Ai, \enquote{Highly focused high-frequency
		travelling surface acoustic waves {(SAW)} for rapid particle sorting,} Lab.
	Chip \textbf{16}, 471--479 (2016).
	
	\bibitem{hendrickson_integrated_2014}
	S.~M. Hendrickson, A.~C. Foster, R.~M. Camacho, and B.~D. Clader,
	\enquote{Integrated nonlinear photonics: emerging applications and ongoing
		challenges,} J. Opt. Soc. Am. B \textbf{31}, 3193--3203 (2014).
	
	\bibitem{shi_acousto-optic_2017}
	Y.~Shi, A.~Cerjan, and S.~Fan, \enquote{Acousto-optic finite-difference
		frequency-domain algorithm for first-principles simulations of on-chip
		acousto-optic devices,} {APL} Photonics \textbf{2}, 020801 (2017).
	
\end{thebibliography}
\end{document}